\documentclass[preprint,3p,times]{elsarticle}

\usepackage{ecrc}

\volume{00}

\firstpage{1}

\journalname{Physica A}

\runauth{Li et al.}

\jid{Physica A}

\jnltitlelogo{}

\usepackage{amssymb}
\usepackage{amsthm}

\usepackage{textcomp}
\usepackage{amsmath}

\usepackage{mathrsfs}
\usepackage{mathptmx}

\usepackage{amsfonts} 
\usepackage{thmtools}
\usepackage{remreset}
\usepackage{multirow}

\usepackage{tikz}

\usepackage{stfloats}

\usepackage{graphicx}

\usepackage{booktabs}

\usepackage{algorithm}

\usepackage{float}

\usepackage{algpseudocode}
\usepackage{boondox-cal}

\usepackage{lipsum}

\usepackage{xcolor}
\usepackage[labelfont={color=black}]{caption}
\usepackage[normalem]{ulem}
\usepackage{cancel}
\usepackage{anyfontsize}
\usepackage{subcaption}
\DeclareUnicodeCharacter{0301}{\'{e}}
\DeclareUnicodeCharacter{0308}{\'{o}}

\newtheorem{theorem}{Theorem}
\newtheorem{corollary}{Corollary}
\newtheorem{lemma}{Lemma}

\newtheorem{remark}{Remark}

\usepackage{url}
\biboptions{sort&compress}

\usepackage[figuresright]{rotating}
\usepackage{bm}
\allowdisplaybreaks
\begin{document}

\begin{frontmatter}



\dochead{}
%% e.g. \dochead{17th International Conference on Dynamical Processes in Excited States of Solids}

\title{Sufficient Control of Complex Networks}


\author[inst1]{Xiang Li}
\ead{xiang002@e.ntu.edu.sg}
\address[inst1]{School
of Electrical and Electronic Engineering, Nanyang Technological University, Singapore 639798, Singapore.}

\author[inst2,inst3,inst4]{Guoqi Li}
\ead{guoqi.li@ia.ac.cn}
\address[inst2]{Institute of Automation, Chinese Academy of Sciences, Beijing 100045, China.}
\address[inst3]{School of Artificial Intelligence, University of Chinese Academy of Sciences, Beijing 101408, China.}
\address[inst4]{Peng Cheng Laboratory, Shenzhen 518055, China.}

\author[inst5]{Leitao Gao}
\ead{gaoleitao123@stu.xjtu.edu.cn}
\address[inst5]{The $54^{\text{th}}$ Research Institute of CETC, Shijiazhuang 050081, China.}

\author[inst6]{Beibei Li}
\ead{eackot@ntu.edu.sg}
\ead{libeibei@scu.edu.cn}
\address[inst6]{School of Cyber Science and Engineering, Sichuan University, Chengdu 610065, China.}

\author[inst1]{Gaoxi Xiao\corref{cor1}}
\ead{egxxiao@ntu.edu.sg}
\cortext[cor1]{Corresponding author}

\begin{abstract}
In this paper, we propose to study {\it sufficient control} of complex networks, which is to control a sufficiently large portion of the network, where only the quantity of controllable nodes matters. To the best of our knowledge, this is the first time that such a problem is investigated. We prove that the sufficient controllability problem can be converted into a minimum-cost flow problem, for which an algorithm with polynomial complexity can be devised. Further, we study the problem of minimum-cost sufficient control, which is to drive a sufficiently large subset of the network nodes to any predefined state with the minimum cost using a given number of controllers. The problem is NP-hard. We propose an ``extended $L_{\mathrm{0}}$-norm-constraint-based Projected Gradient Method" (eLPGM) algorithm, which may achieve suboptimal solutions for the problems at small or medium sizes. To tackle the large-scale problems, we propose to convert the control problem into a graph problem and devise an efficient low-complexity ``Evenly Divided Control Paths" (EDCP) algorithm to tackle the graph problem. Simulation results on both synthetic and real-life networks are provided, demonstrating the satisfactory performance of the proposed methods.
\end{abstract}


\end{frontmatter}

% \linenumbers


\section{Introduction}
\label{sec:introduction}

Complex system research has grown steadily over the past two decades, with broad applications in many fields such as brain intelligence, medical science, social science, biology, and economics \cite{Watts1998, Strogatz2001, BARABASI2003}. Many of these complex systems can be modeled as static or dynamic networks, which stimulates the emergence and booming developments of research on complex networks. Two fundamental issues related to the control of complex networks have been extensively studied. They include (i) the structural controllability problem \cite{Lin1974, Liu2011, Dion2003}, which aims at finding the minimum number of driver nodes connected to external controllers for ensuring that the network theoretically speaking can be driven to any state given sufficient time and energy inputs; and (ii) the minimum-cost control problem \cite{Yan2012, Li2015, Tzoumas2016, Summers2016}, which aims at minimizing the cost for driving the system to any predefined state with a given number of controllers. Another problem closely related to the structural control problem is the target control problem \cite{Gao2014, Klickstein2017, Li2020}, which is to control a specific given set of target nodes rather than the whole network.

Various algorithms have been proposed to tackle the problems related to the structural controllability of complex networks. Maximum matching (MM), a classic concept in graph theory, has been successfully and efficiently used to guarantee the structural controllability of the whole network by allocating the minimum number of external control sources with input signals to the network nodes {\cite{Liu2011, Dion2003}}. However, the MM algorithm may not be easily applied to large-scale networks when global topological information is not available. Li {\it et al.} proposed a local-game matching (LM) algorithm to solve the structural controllability problem using only the local network topology information \cite{Li2018}. Other results include \cite{Ruths2014, Liu2016}, which studied control profile, approximate solution, and so on.

The minimum-cost control problem is arguably an even more important problem. Studies have been carried out on this problem in recent years, generating some interesting results \cite{Yan2012, Li2015, Ding2017, Shirin2017, Yan2015, Wang2017, Lindmark2018}. Li {\it et al.} devised the matrix derivative projected gradient descent method and its extensions to iteratively search for the optimal placement of control sources for minimizing control cost \cite{Li2015, Li2018}. 
Projected Gradient Method (PGM), Orthonormal-constraint-based Projected Gradient Method (OPGM), and other extensions have been proposed {\cite{Li2015, Li2018}}. It was found that OPGM slightly outperforms PGM \cite{Li2015} since the driver nodes selected by OPGM are more concentrated on those nodes which divide an elementary dilation equally for achieving a lower energy cost. Inspired by such observations, a low complexity algorithm termed ``Minimizing the Longest Control Path" (MLCP) was proposed \cite{Li2018}, aiming at controlling a complex network at low cost without the complete knowledge of its global topology. Lindmark {\it et al.} proposed a set of rules and strategies for selecting the driver nodes to minimize control cost based on the (weighted) topology of the network \cite{Lindmark2018}. However, as claimed by the authors, the strategy they proposed is inspired by topological considerations but still lacks theoretical justification.

The target controllability problem is to allocate a minimum number of external inputs for a given network such that the state of the prescribed target subset can be driven to any state in finite time with properly designed input. The problem has attracted increasing research interest in recent years \cite{Gao2014, Li2020, VanWaarde2017}. In \cite{Li2020}, Maximum Flow based Target Path-cover (MFTP) algorithm was proposed to tackle the target controllability problem. The objective of the minimum-cost target control problem is to minimize the cost of driving the target set to any predefined state with a given number of controllers. In \cite{Gao2018}, the ``$L_{\mathrm{0}}$-norm-constraint-based Projected Gradient Method" (LPGM) was proposed to solve the minimum-cost target control problem. There are also some other interesting related results. In \cite{Chen2020}, instead of assuming that there exists a predefined target node set, Chen {\it et al.} investigated the case where a number of controllers have been connected to the network, and the objective is to select a set of target nodes such that the control cost is minimized. In \cite{Klickstein2017}, Klickstein {\it et al.} derived the upper bound of the energy required to optimally control a portion of a network. They showed that this upper bound of energy increases exponentially with the size of target nodes given a fixed number of controllers.

We propose to study the {\it sufficient control} problem of which the objective is to control a large enough portion of network nodes, where the size of the controlled portion, or equivalently the number of controlled network nodes, is not smaller than a certain preset threshold. Such control arguably has wide applications in real life \cite{Rainer2002, Masuda2015, Liu2014, Joseph2021, Anderson1992, Sayeed2019}. For example, to win a two-candidate election, it may be sufficient to gain support from slightly more than 50\% of all the voters \cite{Masuda2015, Joseph2021}; to arrest the spreading of an enormously infectious disease with herd immunization, 80\% or more of the population may need to be persuaded to take vaccination \cite{Anderson1992}; and the adversary who can control 51\% or more of network nodes may carry out a highly damaging attack on the blockchain \cite{Sayeed2019}, etc. To the best of our knowledge, this is the first time that such a problem is investigated. In this paper, we mainly study two fundamental issues of the sufficient control problem, (i) \textit{sufficient controllability}, which is to allocate a minimum number of external sources to guarantee the structural controllability of a sufficiently large number of network nodes; and (ii) \textit{minimum-cost sufficient control problem}, which is to minimize the cost of controlling these nodes. The main contributions of our work are summarized as follows:

\begin{itemize}
	\item We prove that the sufficient controllability problem is essentially a maximum controllable subset problem, which can be further converted into a path cover problem. The path cover problem is to locate a minimum set of directed paths denoted as $\mathcal{P}$ and cycles denoted as $\mathcal{C}$ to cover a sufficiently large portion of network nodes.
	\item An algorithm termed ``Minimum-Cost Flow based Path-cover" (MCFP) is proposed to solve the path cover problem with polynomial complexity. We also apply the MCFP method on some synthetic and real-life networks and observe the sufficient controllability of these networks. 
	\item We formulate the minimum-cost sufficient control problem which is NP-hard. To tackle this problem, we propose an ``extended $L_{\mathrm{0}}$-norm-constraint-based Projected Gradient Method" (eLPGM) which works efficiently on small- or medium-sized networks. 
	\item For large scale networks, an ``Evenly Divided Control Path" (EDCP) algorithm is proposed to tackle the minimum-cost sufficient problem at low complexity. Simulation results on synthetic and real-life networks evidently demonstrate the effectiveness of both the eLPGM and EDCP algorithms. 
\end{itemize}

The remaining part of this article is organized as follows. In Section \ref{sec:preliminaries}, the graphic theoretical explanation for network controllability is introduced. In Section \ref{sec:Sufficient Controllability of Directed Networks}, the sufficient controllability problem is formulated. The algorithm named MCFP is proposed. Simulation results on both synthetic and real-life networks are provided. In Section \ref{sec:Minimum-cost Sufficient Control Problem}, the minimum-cost sufficient control problem is formulated. The eLPGM and EDCP algorithms are proposed. The performances of the proposed algorithms are evaluated. Section \ref{sec:Conclusion} concludes the article.

% Throughout this paper, we use plain lowercase letters to denote scalar or scalar-valued functions, bold lowercase letters to denote vector or vector-valued functions, plain uppercase letters to denote matrices, and bold uppercase letters to denote sets.}

% \section{Related Works}
% \label{sec:Related Works}

\section{Graphic Theoretical Explanation for Network Controllability}
\label{sec:preliminaries}
This section provides preliminaries, including structural controllability, target controllability, and network flow. Although most real-world systems are driven by nonlinear processes, understanding the impacts of topological characteristics on linear control is naturally a prerequisite for tackling nonlinear control problems. As that in many of the references listed earlier, in this report, we carry out our study based on the canonical linear time-invariant (LTI) model.

\subsection{Structural Controllability}
Consider the following system with linear dynamics
\begin{equation} 
	\dot{\mathbf{x}}(t) = A\mathbf{x}(t) + B\mathbf{u}(t)
	\label{eq.system}
\end{equation}
where $\mathbf{x}(t) \in \mathbb{R}^{N \times 1}$ is the state vector of {\it N} nodes at time \textsl{t} with an initial state $\textbf{x}(0)$, $\mathbf{u}(t) \in \mathbb{R}^{M \times 1}$ is the time-dependent external control input vector of {\it M} external control sources in which the same input $u_{i}(t)$ may connect to multiple nodes. The system matrix $A = (a_{ij}) \in \mathbb{R}^{N \times N}$ is the transpose of the adjacency matrix of the network, i.e., the matrix element $a_{ij} \ne 0$ if there is a link connecting node \textsl{j} to node \textsl{i}; and $a_{ij} = 0$ otherwise. The matrix $B =(b_{ij}) \in \{0, 1\}^{N \times M}$ is the input matrix where $b_{im}$ is one when controller \textsl{m} is connected to node \textsl{i} and zero otherwise. Hereafter, we will use $(A,B)$ to refer to the system (\ref{eq.system}). In a classical article, Lin {\cite{Lin1974}} gives the following conclusion about the structural controllability.

\begin{lemma}
\label{lemma_structural_controllability}
{\cite{Lin1974}}
A linear system $(A, B)$ is structurally controllable if and only if there is no inaccessible node or no dilation in the digraph $D({V}, {E})$.
\end{lemma}

\begin{remark}
The graph $D({V}, {E})$ is mapped from an LTI system $(A, B)$ through function $G: (A, B) \rightarrow D(V, E)$. This mapping can be denoted as $G(A, B) = D(V, E)$, where the vertex set ${V} = V_{\text{A}} \cup V_{\text{B}}$ and the edge set ${E} = E_{\text{A}} \cup E_{\text{B}}$. Here $V_{\text{A}}$ and $E_{\text{A}}$ represent the vertex set and the edge set in the original network, respectively. $V_{\text{B}}$ represents $M$ external controllers or input vertices and $E_{\text{B}}$ represents the edge set from controllers to the original network vertex set based on $B$. A node $v_{i}$ in $G(A, B)$ is called inaccessible if and only if it is not possible to be reached from any input vertices in $V_{B}$. The digraph $G(A, B)$ contains a dilation if and only if there is a subset $Z \subseteq V_{\text{A}}$ such that $|V_{\text{nbhd}}(Z)| < |Z|$. Here the neighbourhood set $V_{\text{nbhd}}(Z)$ = $\{v_j: (v_j, v_i) \in E, v_i \in Z\}$ includes all nodes $v_j$ where there is a directed edge from $v_{j}$ to any node in $Z$, and $|\cdot|$ denotes the set cardinality.
\end{remark}

\begin{lemma}
\label{lemma_cacti}
\cite{Liu2011}
A linear system is structurally controllable if and only if there exists a vertex-disjoint union of cacti in the digraph $G(A, B)$ that covers all the vertices in $V_{\text{A}}$. By denoting $B = [b_1, \cdots, b_m]$ where $b_i$ is the $i^{\text{th}}$ column of $B$, this criterion is fulfilled if and only if for any
$i, j \in \{1,2,\cdots,M\}$ and $i \neq j$,
	
a) $G(A, b_{i})$ contains a cactus.
	
b) The cacti contained in $G(A, b_i)$ and $G(A, b_j)$ are vertex-disjoint. 
	
c) The cacti contained in $G(A, b_{1}), G(A, b_{2}),\cdots,G(A, b_{M})$ cover all vertices in $V_{\text{A}}$.
\end{lemma}

\begin{remark}
\label{rm_cacti} A path on a directed graph is a sequence of adjacent nodes where each node appears only once, while a cycle is a sequence of adjacent nodes where the initial and final nodes coincide. A stem $\mathcal{l}$ is a path starting from an input vertex. A cycle and an additional edge pointing to a vertex of the cycle form a bud $\mathcal{o}$. This additional edge is called the distinguished edge of the bud. A cactus $\mathcal{t}$ consists of one stem and $k$ buds, i.e., $\mathcal{t} = \mathcal{l} \cup \mathcal{o}_1 \cup \cdots \cup \mathcal{o}_k$. For any bud $\mathcal{o}_i$, $(i \in [k])$, the starting vertex of its distinguished edge is the only node belonging to $\mathcal{o}_i $ and $\mathcal{l} \cup \mathcal{o}_1 \cup \cdots \cup \mathcal{o}_{i-1} $ simultaneously \cite{Liu2011}. The originating vertex of the stem is called the root of the stem, which is also the root of the cactus. Note that a stem is a cactus with zero buds. A node is called covered by a path/cycle/cactus if the path/cycle/cactus includes the node.
\end{remark}

\subsection{Target Controllability}
Given the target set $S = \{ v_{s_1},v_{s_2},\cdots,v_{s_{|S|}}\} \subseteq V_{\text{A}}$ where $s_i$ represents the index of the $i^{\text{th}}$ target node, we have the following LTI system
\begin{equation}
	\begin{aligned}
		&\dot{\mathbf{x}}(t)= A\mathbf{x}(t) + B\mathbf{u}(t)\\
		&\mathbf{y}(t) = C \mathbf{x}(t)
	\end{aligned}
	\label{eq.LTI_traget_controllability}
\end{equation}
where $\mathbf{y}(t) \in \mathbb{R}^{|S| \times 1} (|S| \leq N)$ denotes the system output. The output matrix $C =(c_{ij}) =[I_{s_1},\cdots,I_{s_i},\cdots,I_{s_{|S|}}]^\text{T} \in \mathbb{R}^{|S| \times N}$ consists of $|S|$ rows of an $N\times N$ identity matrix, where $I_{s_i}$ denotes the $s_i^{\text{th}}$ row. In other words, the entry $c_{is_{i}} = 1$ if $v_{s_i} \in S$; and $c_{is_{i}} = 0$ otherwise. Target controllability is to determine the minimum number of external control sources required such that the state of the target set $S$ can be driven to any state in finite time with a properly designed input $\mathbf{u}(t)$. The system (\ref{eq.LTI_traget_controllability}) is said to be target/output controllable if and only if $\operatorname{rank}[CB, CAB,\cdots, CA^{N-1}B] = |S|$ for a determined input matrix $B$, pre-defined $A$ and the target set $S$ {\cite{Gao2014}}. The graphic theoretical explanation for target controllability is given as follows. 

\begin{lemma}
\label{lemma_target_controllability}
\cite{Blackhall2010}
The system (\ref{eq.LTI_traget_controllability}) is target controllable if every vertex in $S$ can be covered by a union of cacti structure contained in the digraph $G(A, B)$. 
\end{lemma}

% By permuting the node index in $V_{\text{A}}$ such that the first $|S|$ nodes become the target nodes, then $\mathbf{x}_1 = [x_1,\cdots,x_{|S|}]^\text{T}$ and $\mathbf{x}_2 = [x_{|S|+1},\cdots,x_N]^\text{T}$ denote the state of the target set $S$ and non-target set $V_{\text{A}} \setminus S$, respectively. We rewrite (\ref{eq.LTI_traget_controllability}) as
% \begin{equation}
% 	\begin{aligned}
% 		\begin{bmatrix}
% 			\dot{\mathbf{x}}_1(t)\\
% 			\dot{\mathbf{x}}_2(t)
% 		\end{bmatrix}
% 		&=
% 		\begin{bmatrix}
% 			A_{11}&A_{12}\\
% 			A_{21}&A_{22}
% 		\end{bmatrix}
% 		\begin{bmatrix}
% 			\mathbf{x}_1(t)\\
% 			\mathbf{x}_2(t)
% 		\end{bmatrix}+
% 		\begin{bmatrix}
% 			B_1\\
% 			B_2
% 		\end{bmatrix} \mathbf{u}(t)
% 	\end{aligned}
% 	\label{eq.targetcontrollability}
% \end{equation} 
% where $A_{11}$ and $A_{22}$ are the transpose of the adjacency matrix of the target node set $S$, and the transpose of the adjacency matrix of the non-target nodes in the set $V \setminus S$, respectively. $A_{12}$ and $A_{21}$ represent the connections between $S$ and $V \setminus S$. $B_{1}$ and $B_{2}$ represent the input matrices connected to $S$ and $V \setminus S$, respectively.
% We have the following lemma.

\subsection{Network Flow}
A flow network $D(V, E, b, c, s)$ is a directed graph with vertex set $V$, edge set $E$, a nonnegative capacity function $b: E \rightarrow \mathbb{N}$, a cost function $c: E \rightarrow \mathbb{Z}$, and an integral supply function $s: V \rightarrow \mathbb{Z}$ satisfies $\sum_{v \in V}s(v) = 0$ \cite{Gross2003}.

% \textbf{\textsl{Single-source single-sink network (or $s-t$ flow network)}} \cite{Gross2003}: An $s-t$ flow network is a flow network that contains two distinguished vertices source $s$ and sink $t$ such that $s(v) = 0$ for all intermediate vertices $v \in V / \{s, t\}$ and $s(s) = -s(t) > 0$. 

A (feasible) network flow is a function \textsl{f}: $E \rightarrow \mathbb{N}$ subject to the following constraints \cite{Gross2003}:
		
a) \textsl{capacity constraints:} $f{(v_i, v_j)} \leq b{(v_i, v_j)}$ for all $(v_i, v_j) \in E$. 

b) 	\textsl	{nonnegativity constraints}: $f{(v_i, v_j)} \geq 0$ for all $(v_i, v_j) \in E$. 

c) \textsl{flow conservation constraints:} $ \sum_{(v_i, v_j)\in E} f{(v_i, v_j)} - \sum_{(v_k, v_i)\in E}f{(v_k, v_i)}= s(v_i) $ for each $v_i \in V$.
 
A minimum-cost flow $f^*$ is a flow with the minimum total flow cost, i.e., $f^* = \operatorname*{argmin}_{f} \sum_{{(v_i, v_j)} \in E}f{(v_i, v_j)} \cdot c{(v_i, v_j)}$ \cite{Gross2003}.
		
\section{Sufficient Controllability of Directed Networks}
\label{sec:Sufficient Controllability of Directed Networks}
In this part, we study the sufficient controllability problem. First, we shall show that the sufficient controllability problem is essentially a maximum controllable subset problem, which can be further converted into a path cover problem. Then an algorithm termed ``Minimum-Cost Flow based Path-cover" (MCFP) is proposed to solve the path cover problem at low complexity. We rigorously prove that this method provides the minimum number of control sources needed to achieve the sufficient controllability on arbitrary directed networks. Finally, we apply the MCFP method to a series of synthetic and real-life networks and observe the sufficient controllability of these networks. 

\subsection{Sufficient Controllability Problem Formulation}
Sufficient controllability problem is to allocate connections between a minimum number of external control inputs and the network to ensure that a given sufficiently large number of nodes are controllable. The sufficient controllability problem can be formulated as 

% \begin{problem}
\begin{equation}
	\begin{aligned}
		& \min_{B, C, M} M \\
		\operatorname{s.t.} \quad 
% & \dot{\mathbf{x}}(t)= A\mathbf{x}(t) + B\mathbf{u}(t)\\
		% &\mathbf{y}(t) = C \mathbf{x}(t)\\
&B \in \{0, 1\}^{N \times M},\\
& C \in \{0, 1\}^{T \times N}, \\
&\operatorname{rank}(C) \geq T,\\
		&\operatorname{rank}[CB, CAB,\cdots, CA^{N-1}B] \geq T,\\
	\end{aligned}
	\label{eq.sufficient_controllability}
\end{equation}
% \label{problem.sufficient_controllability}
% \end{problem}
where $T$ is the number of nodes that are required to be controlled and $M$ is the number of controllers. It is worth noting that the output matrix $C$ is composed of rows from the identity matrix, i.e.,
$C = (c_{ij}) =[I_{r_1},\cdots,I_{r_i},\cdots,I_{r_{T}}]^\text{T}$ where $I_{r_i}$ denotes the $r_i^{\text{th}}$ row of an $N\times N$ identity matrix. The controllable node set thus can be represented as $R =\{ v_{r_1},v_{r_2},\cdots, v_{r_{T}}\} \subseteq V_{\text{A}}$.

It can be easily seen that when $T = N$, the sufficient controllability problem is reduced to the classical structural controllability problem, which can be well solved by employing the maximum matching algorithm \cite{Liu2011}. If there is a prescribed target set $S$ which is a subset of $V$, the controllability of $S$, known as the target controllability problem, can be well solved by implementing the Maximum Flow based Target
Path-cover (MFTP) algorithm \cite{Li2020}. However, when there is no specific controllable node set and the only concern is the quantity of the controllable nodes, neither the MM algorithm nor the MFTP algorithm can be applied directly, and the problem becomes fundamentally different.

\subsubsection{Converting Sufficient Controllability Problem into the Maximum Controllable Subset Problem}
Based on Lemmas \ref{lemma_cacti} and \ref{lemma_target_controllability}, sufficient controllability is to identify the minimum number of vertex-disjoint cacti in $D(V, E)$ that can cover at least $T$ nodes belonging to $V_{\text{A}}$. The minimum number of external inputs needed for sufficient control is equal to the minimum number of vertex-disjoint cacti. An approach to solve this problem is to gradually increase the number of vertex-disjoint cacti and find the maximum number of nodes that can be covered by these cacti. This process can be repeated iteratively until a sufficient portion of nodes are covered. 
We refer to finding the maximum number of nodes that can be covered by a given number of vertex-disjoint cacti as the maximum controllable subset problem.

\begin{figure}[h!]
	\centering
	\begin{subfigure}{0.48\textwidth}
		\includegraphics[width=\textwidth]{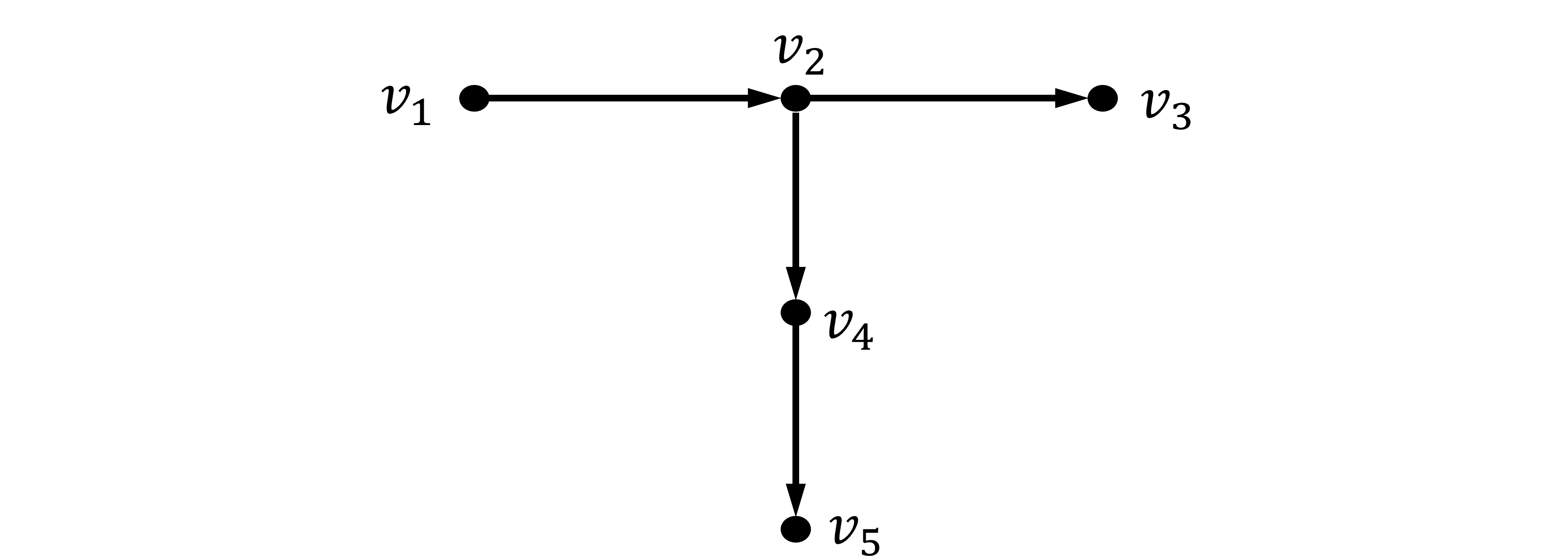}
		\caption{}
		\label{Fig.path_cover_minimum_cost_flow.sub.1}
	\end{subfigure}
% \hspace{1cm}
	\begin{subfigure}{0.48\textwidth}
		\includegraphics[width=\textwidth]{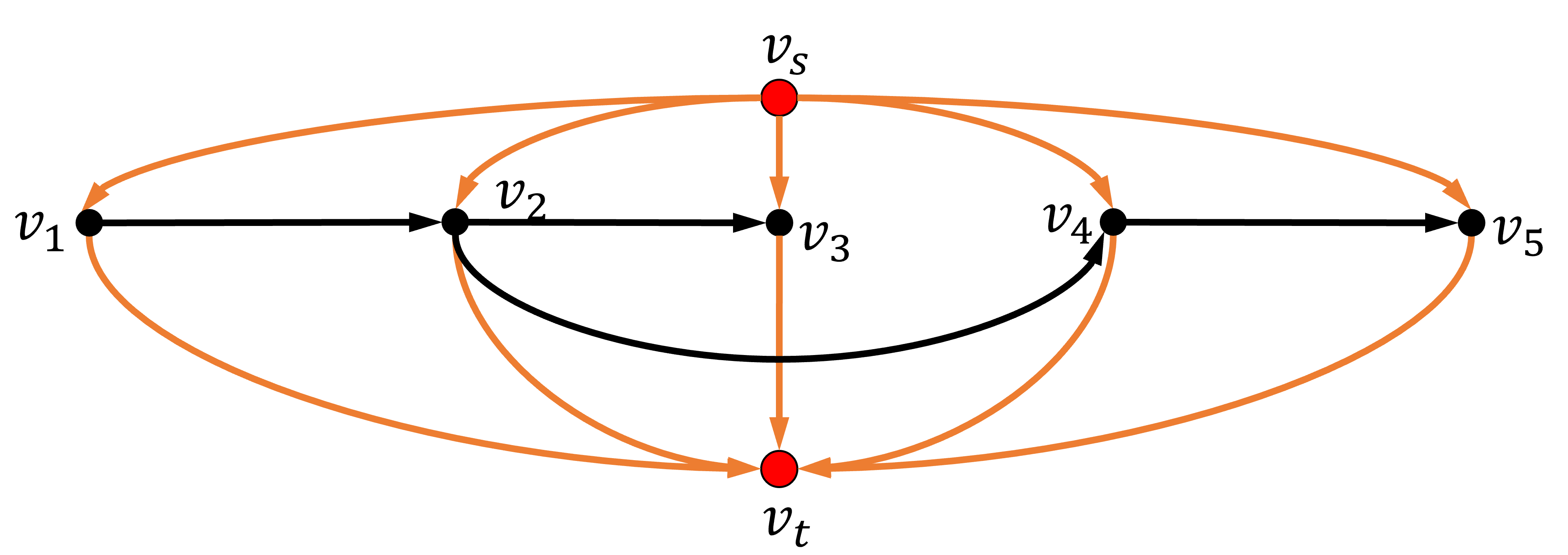}
		\caption{}
		\label{Fig.path_cover_minimum_cost_flow.sub.2}
	\end{subfigure}
\hspace{0cm}
	\begin{subfigure}{0.48\textwidth}
		\includegraphics[width=\textwidth]{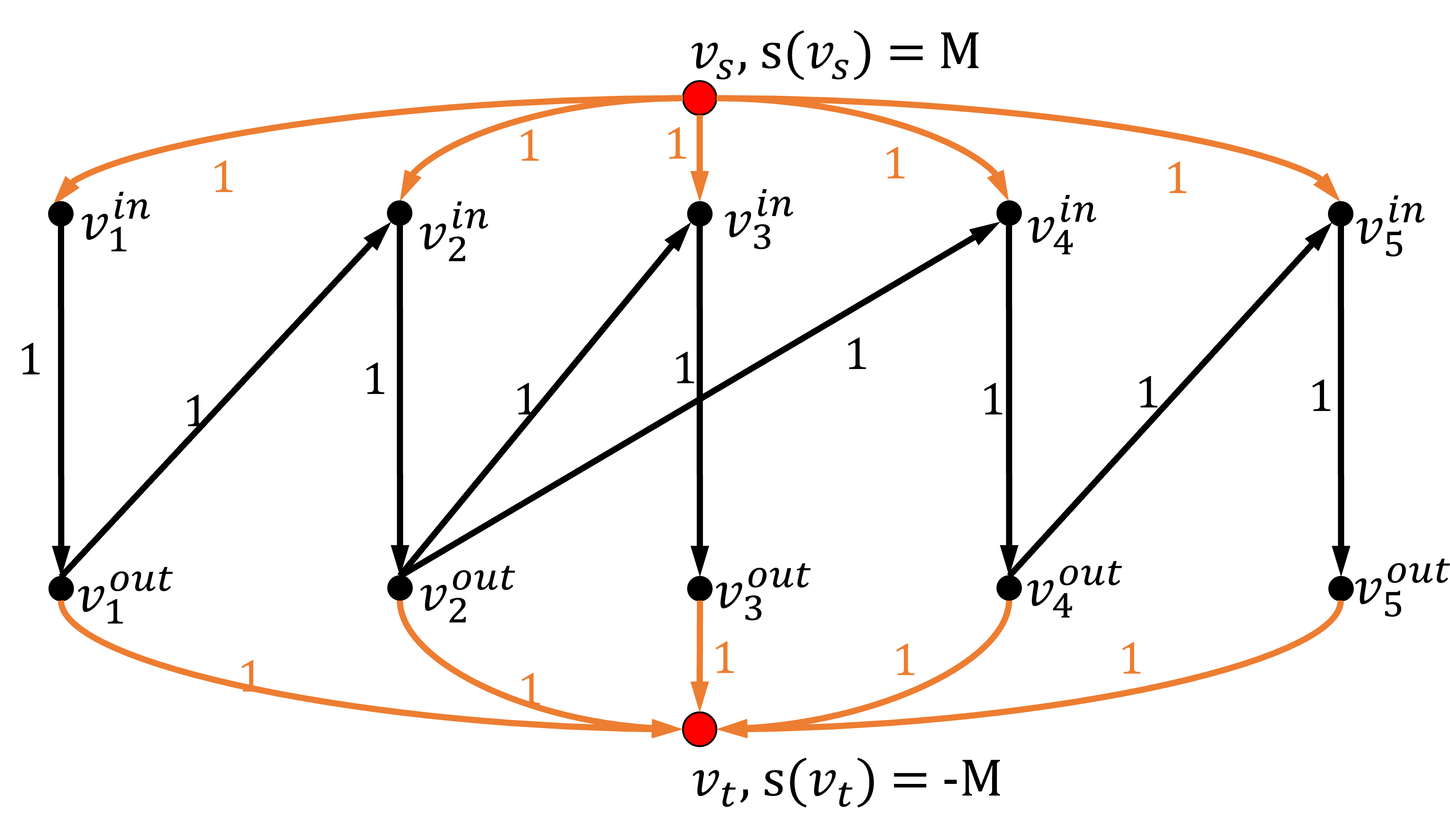}
		\caption{}
		\label{Fig.path_cover_minimum_cost_flow.sub.3}
	\end{subfigure}
% \hspace{-0.3cm}
	\begin{subfigure}{0.48\linewidth}
		\includegraphics[width=\textwidth]{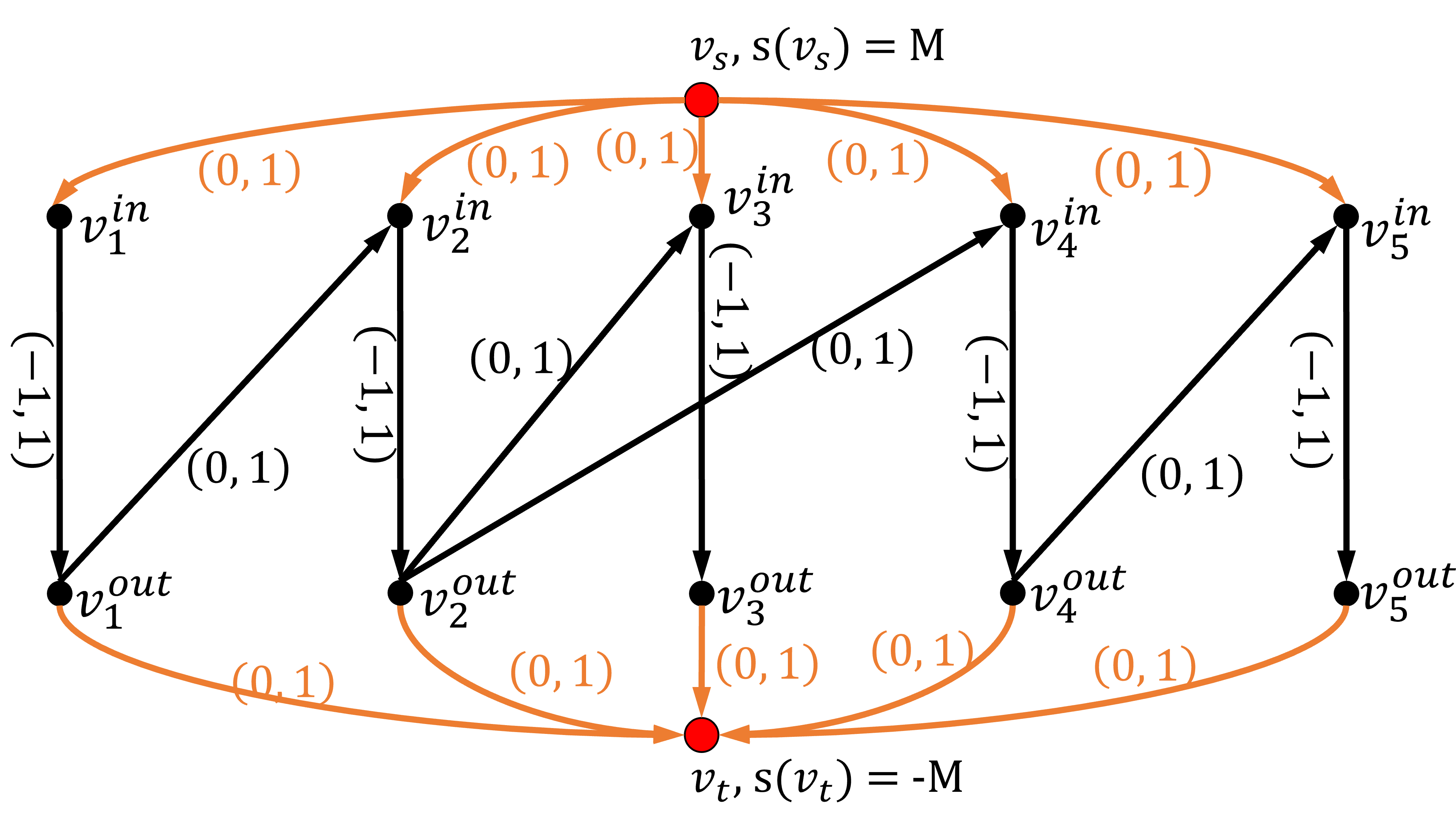}
		\caption{}
		\label{Fig.path_cover_minimum_cost_flow.sub.4}
	\end{subfigure}
	
	\caption[Graph transformation]
	{Transfer a network $D(V_\text{A}, E_\text{A})$ into a flow network $D(V', E', b, c, s)$. (a) The network $D(V_\text{A}, E_\text{A})$ with 5 vertices and 3 edges. (b) Add two distinguished vertices source $v_s$ and sink $v_t$, arcs from $v_s$ to other intermediate vertices and from other intermediate vertices to $v_t$. 
		(c) Split each intermediate vertex $v_i \in V_\text{A}$ into two vertices, $v^{\text{in}}_i$ and $v^{\text{out}}_i$, and add an edge from $v^{\text{in}}_i$ to $v^{\text{out}}_i$. The original edges entering(exiting) $v_i$ now enter $v^{\text{in}}_i$ (exit $v^{\text{out}}_i$).
The capacity of every edge is set to be one. The numerical value shown next to each edge represents its capacity. The supply of the two distinguished vertices $v_s$ and $v_t$ are set as $s(v_s) = -s(v_t) = M$, while the supplies of all other intermediate vertices are set to zero. 
		(d) By setting the cost on each edge from $v_i^{\text{in}}$ to $v_i^{\text{out}}$ to be $-1$, and the cost on the rest of the edges to be zero, the network $D(V_\text{A}, E_\text{A})$ is converted into a flow network $D(V', E', b, c, s)$. The numerical pairs $(c, b)$ shown next to each edge are used to represent the cost and capacity of the edge.}
	\label{Fig.path_cover_minimum_cost_flow}
\end{figure}

\subsubsection{Converting the Maximum Controllable Subset Problem into Path Cover Problem}

The maximum controllable subset problem can be converted into a path cover problem in $D(V_{\text{A}}, E_{\text{A}})$. In particular, 
we denote a set of directed vertex-disjoint paths by $\mathcal{P} = \{ \mathcal{p}_1, \mathcal{p}_2, \cdots, \mathcal{p}_{|\mathcal{P}|}: \mathcal{p}_i \cap \mathcal{p}_j = \emptyset, \;\ i,j \in [|\mathcal{P}|], \forall i \neq j\}$, where $\mathcal{p}_i$ represents the $i^{\text{th}}$ path in $\mathcal{P}$ and the cardinality $|\mathcal{p}_i|$ is the number of nodes on the path. $[n]$ represents the natural number set including all natural numbers from 1 to $n$, i.e., $[n] = \{1,2,\cdot,n\}$, where $n \in \mathbb{N}^{+}$. 
Similarly, we denote a set of directed vertex-disjoint cycles by ${\mathcal{C}} =\{ \mathcal{c}_1, \mathcal{c}_2,\cdots, \mathcal{c}_{|\mathcal{C}|}: \mathcal{c}_i \cap \mathcal{c}_j = \emptyset, \;\ i,j \in [|\mathcal{C}|], \forall i \neq j\}$, where $|\mathcal{c}_i|$ denotes the number of nodes on the $i^{\text{th}}$ cycle $\mathcal{c}_i$. Moreover, for the set of vertex-disjoint paths and cycles $\mathcal{P} \cup {\mathcal{C}}$, we use $\operatorname{cover}(\mathcal{P} \cup {\mathcal{C}})$ to represent a node set consisting of all network nodes covered by paths in $\mathcal{P}$ or cycles in $\mathcal{C}$. The cardinality $|\operatorname{cover}(\mathcal{P} \cup {\mathcal{C}})| = \Sigma^{|\mathcal{P}|}_{k=1} |\mathcal{p}_k| + \Sigma^{|\mathcal{C}|}_{l= 1}|{\mathcal{c}}_l|$ is the number of covered nodes. These covered nodes refer to the structural controllable nodes.
The maximum controllable subset problem with $M$ control sources thus can be formulated as the following path cover problem,

\begin{equation}
	\begin{aligned}
		&\max_{\mathcal{P}, \mathcal{C}} &&|\operatorname{cover}(\mathcal{P} \cup {\mathcal{C}})|\\
		% &\text{s.t.} &&|\textsl{cover}(\mathcal{P} \cup {\mathcal{C}})| = \sum_{k=1}^{|\mathcal{P}|}|\mathcal{p}_k| + \sum_{l=1}^{|{\mathcal{C}}|}|{\mathcal{c}}_l|,\\ 
		&\text{s.t.} && |\mathcal{P}| = M \geq 1, \\
 &&&\mathcal{P} = \{ \mathcal{p}_1, \mathcal{p}_2,\cdots, \mathcal{p}_{|\mathcal{P}|}: \mathcal{p}_i \cap \mathcal{p}_j = \emptyset, \;\ i,j \in [|\mathcal{P}|], \forall i \neq j\},\\
 &&&{\mathcal{C}} =\{ \mathcal{c}_1, \mathcal{c}_2,\cdots, \mathcal{c}_{|\mathcal{C}|}: \mathcal{c}_i \cap \mathcal{c}_j = \emptyset, \;\ i,j \in [|\mathcal{C}|], \forall i \neq j\},\\
 &&& \mathcal{p}_i \cap \mathcal{c}_j = \emptyset, \forall \mathcal{p}_i \in \mathcal{P}, \mathcal{c}_j \in \mathcal{C}.\\
	\end{aligned}
	\label{eq.maxpathcover}
\end{equation}

\subsection{Converting the Path Cover Problem into the Minimum-cost Flow Problem}
\label{subsec_Path_Cover_to_Minimum-cost_Flow}
We shall show that the path cover problem formulated in (\ref{eq.maxpathcover}) is essentially a minimum-cost flow problem. Fig. \ref{Fig.path_cover_minimum_cost_flow} illustrates the procedure for transforming a network into a flow network. Given a network $D(V_{\text{A}}, E_{\text{A}})$, we create two additional nodes $v_s$ and $v_t$ to represent the dummy external control sources. For every $v_i \in V_{\text{A}}$, we add two edges $(v_s, v_i)$ and $(v_i, v_t)$. The added vertex set is denoted by $V_{\text{s,t}} = \{v_s, v_t\}$. And the added edge sets are denoted by $E_{\text{s}} = \{(v_s, v_i): v_i \in V_{\text{A}}\}$, and $E_{\text{t}} = \{(v_i, v_t): v_i \in V_{\text{A}}\}$, respectively. The network $D(V_{\text{A}}, E_{\text{A}})$ is converted into $D(V_{\text{A}}\cup V_{\text{s,t}}, E_{\text{A}} \cup E_{\text{s}} \cup E_{\text{t}})$. This process is shown in Figs. \ref{Fig.path_cover_minimum_cost_flow}(a) - \ref{Fig.path_cover_minimum_cost_flow}(b).

To ensure that the paths and cycles are vertex-disjoint, we split every vertex $v_i \in V_{\text{A}}$ into two associate vertices $v_i^{\text{in}}$ and $v_i^{\text{out}}$, and add an edge $(v_i^{\text{in}}, v_i^{\text{out}})$. The edges originally entering (exiting) $v_i$ now enter $v_i^{\text{in}}$ (exit $v_i^{\text{out}}$). 
The edge set is denoted by $E' = E'_{\text{A}} \cup E'_{\text{inner}} \cup E'_{\text{s}} \cup E'_{\text{t}}$, where
$E'_{\text{A}} = \{(v_i^{\text{out}}, v_j^{\text{in}}): (v_i, v_j) \in E_{\text{A}}\}$, $E'_{\text{inner}} = \{(v_i^{\text{in}}, v_i^{\text{out}}): v_i \in V_{\text{A}}\}$, $E'_{\text{s}} = \{(v_s, v_i^{\text{in}}): (v_s, v_i) \in E_{\text{s}}\}$, and $E'_{\text{t}} = \{(v_i^{\text{out}}, v_t): (v_i, v_t) \in E_{\text{t}}\}$. 
The capacity $b(e)$ is set to one for every edge $e \in E'$. The vertex set is denoted by $V' = V_{\text{A}}^{\text{in}} \cup V_{\text{A}}^{\text{out}} \cup V_{\text{s,t}}$, where $V_{\text{A}}^{\text{in}} = \{v_i^{\text{in}}: v_i \in V_{\text{A}}\}$, $V_{\text{A}}^{\text{out}} = \{v_i^{\text{out}}: v_i \in V_{\text{A}}\}$. Since the number of control sources is $M$, we set the supply of $v_s$ to be $M$ and the supply of $v_t$ to be $-M$. The supplies of nodes in $V_{\text{A}}^{\text{in}} \cup V_{\text{A}}^{\text{out}}$ are set to zero. 
The network $D(V_{\text{A}} \cup V_{\text{s,t}}, E_{\text{A}} \cup E_{\text{s}}\cup E_{\text{t}})$ becomes $D(V', E', b, s)$. The resulting network is shown in Fig. \ref{Fig.path_cover_minimum_cost_flow}(c).

Finally, as shown in Fig. \ref{Fig.path_cover_minimum_cost_flow}(d), we assign a cost of $-1$ to each edge in $E'_{\text{A}}$, and a cost of $0$ to all other edges in the network. The network is transformed into $D(V', E', b, c, s)$. 
To solve the path cover problem formulated in (\ref{eq.maxpathcover}), we can use the {\it network simplex} method \cite{Dantzig2003} to find the minimum-cost flow in the network $D(V', E', b, c, s)$. The following theorem and corollary are presented. 

\begin{theorem}
In network $D(V_{\text{A}}, E_{\text{A}})$, the maximum number of structural controllable nodes with $M$ external control sources equals the absolute value of the total cost of the minimum-cost flow in the transformed network $D(V', E', b, c, s)$.
\label{theorem_sufficient_controllability}
\end{theorem}

\begin{proof}
Please refer to \ref{proof_sufficient_controllability}.
\end{proof}

\begin{corollary}
Given a network $D(V_{\text{A}}, E_{\text{A}})$ with $M$ external control sources, the controllable node set containing the maximum number of structurally controllable nodes is $R = \{v_i: f^*(v_i^{\text{in}}, v_i^{\text{out}}) = 1, v_i \in V_{\text{A}}, v_i^{\text{in}} \in V_{\text{A}}^{\text{in}}, v_i^{\text{out}} \in V_{\text{A}}^{\text{out}}\}$, where $f^*$ represents the minimum-cost flow in the transformed network $D(V', E', b, c, s)$. 
\end{corollary}

\subsection{Minimum-Cost Flow based Path-cover Algorithm}
According to Theorem \ref{theorem_sufficient_controllability}, we propose a ``Minimum-Cost Flow based Path-cover" (MCFP) algorithm to solve the path cover problem formulated in (\ref{eq.maxpathcover}). The main idea of the MCFP algorithm is to first transfer a network $D(V_{\text{A}}, E_{\text{A}})$ into a flow network $D(V', E', u, c, s)$, and then obtain the edge set according to the flow paths by applying a minimum-cost flow algorithm on $D(V', E', u, c, s)$. Finally, the control paths and cycles can be allocated. The MCFP algorithm is presented in Algorithm \ref{alg:MCFP Algorithm}. 

\begin{algorithm}[H]
	\begin{small}
		\caption{MCFP Algorithm}
		\begin{algorithmic}[1]
			\renewcommand{\algorithmicrequire}{\textbf{Input:}}
			\renewcommand{\algorithmicensure}{\textbf{Output:}}
			\Require Network $D(V_{\text{A}}, E_{\text{A}})$, the number of controllers $M$
			\Ensure Control path set $\mathcal{P}$,and cycle set $\mathcal{C}$
			%		\\ \textit{Initialisation} :
			%		\\ \textit{LOOP Process}
 \State Add a source node $v_s$ and a sink node $v_t$. Create the node set $V' = V_{\text{A}}^{\text{in}} \cup V_{\text{A}}^{\text{out}} \cup \{v_s, v_t\}$, where $V_{\text{A}}^{\text{in}} = \{v_i^{\text{in}}: v_i \in V_{\text{A}}\}$, $V_{\text{A}}^{\text{out}} = \{v_i^{\text{out}}: v_i \in V_{\text{A}}\}$
 \State Create the edge set $E' = E'_{\text{A}} \cup E'_{\text{inner}} \cup E'_{\text{s}} \cup E'_{\text{t}}$, where
$E'_{\text{A}} = \{(v_i^{\text{out}}, v_j^{\text{in}}): (v_i, v_j) \in E_{\text{A}}\}$, $E'_{\text{inner}} = \{(v_i^{\text{in}}, v_i^{\text{out}}): v_i \in V_{\text{A}}\}$, $E'_{\text{s}} = \{(v_s, v_i^{\text{in}}): (v_s, v_i) \in E_{\text{s}}\}$, and $E'_{\text{t}} = \{(v_i^{\text{out}}, v_t): (v_i, v_t) \in E_{\text{t}}\}$
\State Set $b(e) = 1$ for every $e \in E'$
\State Set $s(v_s) = -s(v_t) = M$; set $s(v) = 0$ for every $v \in V' \setminus \{v_s, v_t\}$
\State Set $c(e) = -1$ for every $e \in E'_{\text{inner}}$; set $c(e) = 0$ for every $e \in E' \setminus E'_{\text{inner}}$
\State Construct the network $D(V', E', b, c, s)$
\State Find the minimum-cost flow $f^*$ in $D(V', E', b, c, s)$
			\State Obtain edge set $E^{f^*} = \{(v_i, v_j):f^*(v_i, v_j) =1\}$, and node set $V^{f^*}$ by extracting all unique nodes from $E^{f^*}$
 \State Find all paths and cycles in network $D(V^{f^*}, E^{f^*})$ using depth-first search (DFS)
   \State Map paths and cycles in $D(V^{f^*}, E^{f^*})$ to the corresponding paths and cycles in $D(V_{\text{A}}, E_{\text{A}})$. The mapped paths compose path set $\mathcal{P}$ and the mapped cycles compose cycle set $\mathcal{C}$
   % \State Initialize $\mathcal{P} = \emptyset$ and ${\mathcal{C}} = \emptyset$
			% \While{$E^{f^*} \neq \emptyset$}		
			% \If {there exists $f^*(v_s, v_1^{\text{in}}) = f^*(v_1^{\text{in}}, v_1^{\text{out}}) = f^*(v_1^{\text{out}}, v_2^{\text{in}}) = \cdots= f^*(v_{\mathcal{p}}^{\text{in}}, v_\mathcal{p}^{\text{out}}) = f^*(v_\mathcal{p}^{\text{out}}, v_t) = 1$,}
			% \State Update $\mathcal{P} = \mathcal{P} \cup \{(v_1, v_2,\cdots,v_l)\}$ and $E^{f^*} = E^{f^*} \setminus \{(v_s, v_1^{\text{in}}), (v_1^{\text{in}}, v_1^{\text{out}}), (v_1^{\text{out}}, v_2^{\text{in}}),\cdots,(v_{\mathcal{p}}^{\text{in}}, v_\mathcal{p}^{\text{out}}), (v_\mathcal{p}^{\text{out}}, v_t) \}$.
			% \EndIf
			% \If {there exists $f(v_1^{\text{in}}, v_1^{\text{out}}) = f(v_1^{\text{out}}, v_2^{\text{in}}) = \cdots= f(v_{\mathcal{c}-1}^{\text{out}}, v_{\mathcal{c}}^{\text{in}}) = f(v_{\mathcal{c}}^{\text{in}}, v_{\mathcal{c}}^{\text{out}}) = f(v_{\mathcal{c}}^{\text{out}}, v_1^{\text{in}}) = 1$,}
			% \State {Update ${\mathcal{C}} = {\mathcal{C}} \cup \{(v_1, v_2,\cdots,v_{\mathcal{c}}\})$ and $E^{f^*} = E^{f^*} \setminus \{(v_1^{\text{in}}, v_1^{\text{out}}), (v_1^{\text{out}}, v_2^{\text{in}}),\cdots,
   % (v_{\mathcal{c}-1}^{\text{out}}, v_{\mathcal{c}}^{\text{in}}),
   % (v_{\mathcal{c}}^{\text{in}}, v_{\mathcal{c}}^{\text{out}}),(v_{\mathcal{c}}^{\text{out}}, v_1^{\text{in}}) \}$}.
			% \EndIf
			% \EndWhile
			%\\ \Return $\mathcal{P}$ and ${\mathcal{C}}$.
		\end{algorithmic} 
	 \label{alg:MCFP Algorithm}
	\end{small}
\end{algorithm}

Lines 1 - 6 of the MCFP algorithm are to artificially transfer a given network $D(V_{\text{A}}, E_{\text{A}})$ into a flow network $D(V', E', b, c, s)$ by the procedures discussed in Section \ref{subsec_Path_Cover_to_Minimum-cost_Flow}. After the graph transfer, line 7 identifies the minimum-cost flow $f^*$ on $D(V', E', b, c, s)$.
Line 8 obtains the edge set $E^{f^*} = \{(v_i, v_j):f^*(v_i, v_j) =1\}$ and its corresponding node set $V^{f^*} = \{v_i: (v_i, v_j)\in E^{f^*} \operatorname{or} \ \ (v_k, v_i)\in E^{f^*}\}$.
As stated in line 9, the paths and cycles on $D(V^{f^*}, E^{f^*})$ can be located by applying depth-first search (DFS) \cite{Gross2003}. Finally, line 10 maps the located paths and cycles to their corresponding paths and cycles in $D(V_{\text{A}}, E_{\text{A}})$. The path set $\mathcal{P}$ and cycle set $\mathcal{C}$ are composed of the mapped paths and cycles, respectively.
Specifically, a path 
\begin{equation}
    (v_s, v_1^{\text{in}}, v_1^{\text{out}}, v_2^{\text{in}}, \cdots, v_{\mathcal{p}-1}^{\text{out}},v_{\mathcal{p}}^{\text{in}}, v_{\mathcal{p}}^{\text{out}}, v_t),
\end{equation}
in $D(V^{f^*}, E^{f^*})$, shall be mapped to a path
\begin{equation}
    (v_1, v_2, \cdots, v_{\mathcal{p}-1},v_{\mathcal{p}}),
\end{equation}
in $D(V_{\text{A}}, E_{\text{A}})$.
Similarly, a cycle 
\begin{equation}
    (v_1^{\text{in}}, v_1^{\text{out}}, v_2^{\text{in}}, \cdots, v_{\mathcal{c}-1}^{\text{out}}, v_{\mathcal{c}}^{\text{in}}, v_{\mathcal{c}}^{\text{out}}, v_1^{\text{in}}),
\end{equation}
in $D(V^{f^*}, E^{f^*})$, shall be mapped to a cycle
\begin{equation}
    (v_1, v_2, \cdots, v_{\mathcal{c}-1}, v_{\mathcal{c}}, v_1),
\end{equation}
in $D(V_{\text{A}}, E_{\text{A}})$. The time complexity of the MCFP algorithm is provided. 

\begin{theorem}
    The time complexity of the MCFP algorithm is $\mathcal{O}\left( |V_{\text{A}}| |E|_{\text{A}}^2 M\right)$.
\end{theorem}
\begin{proof}
The time complexity of each line in the MCFP algorithm is analyzed. The time complexity of lines 1 - 6 of the MCFP algorithm is $\mathcal{O}\left( |V_{\text{A}}| + |E_{\text{A}}| \right)$. 
The minimum-cost flow problem in network $D(V, E)$ can be solved by applying the network simplex method, of which the time complexity is $\mathcal{O}\left( |V| |E|^2 s_{\text{max}} c_{\text{max}} \right)$ \cite{Dantzig2003}. Here $s_{\text{max}}$ denotes the largest absolute value of node supply or edge capacity, i.e.,
\begin{equation}
	s_{\text{max}} = \max\{\max \{|s(v)|: v \in V \}, \: \max\{b(e): e \in E \} \}.
\end{equation}
And $c_{\text{max}}$ denotes the largest absolute value of edge cost:
\begin{equation}
	c_{\text{max}} = \max\{ |c(e)|: e \in E \}. 
\end{equation}
In $D(V', E', b, c, s)$, since $s_{\text{max}} = M$ and $c_{\text{max}} = 1$, the time complexity of line 7 is $\mathcal{O}\left( |V_{\text{A}}| |E_{\text{A}}|^2 M \right)$. 
For line 8, in the worst case, all edges in $E'$ are in $E^{f^*}$. Thus, line 8 has the time complexity of $\mathcal{O}\left( |E_{\text{A}}|\right)$.
And the time complexity of lines 9 - 10 for finding all paths and cycles in $D(V^{f^*}, E^{f^*})$ with DFS is $\mathcal{O}\left( |V^{f^*}||E^{f^*}|\right)$.
Overall, the time complexity of the MCFP algorithm is $\mathcal{O}\left( |V_{\text{A}}| |E|_{\text{A}}^2 M\right)$.
\end{proof}

\subsection{Experimental Results}
\begin{figure}[H]
	\centering
\begin{minipage}{.48\textwidth}
\includegraphics[width=\textwidth]{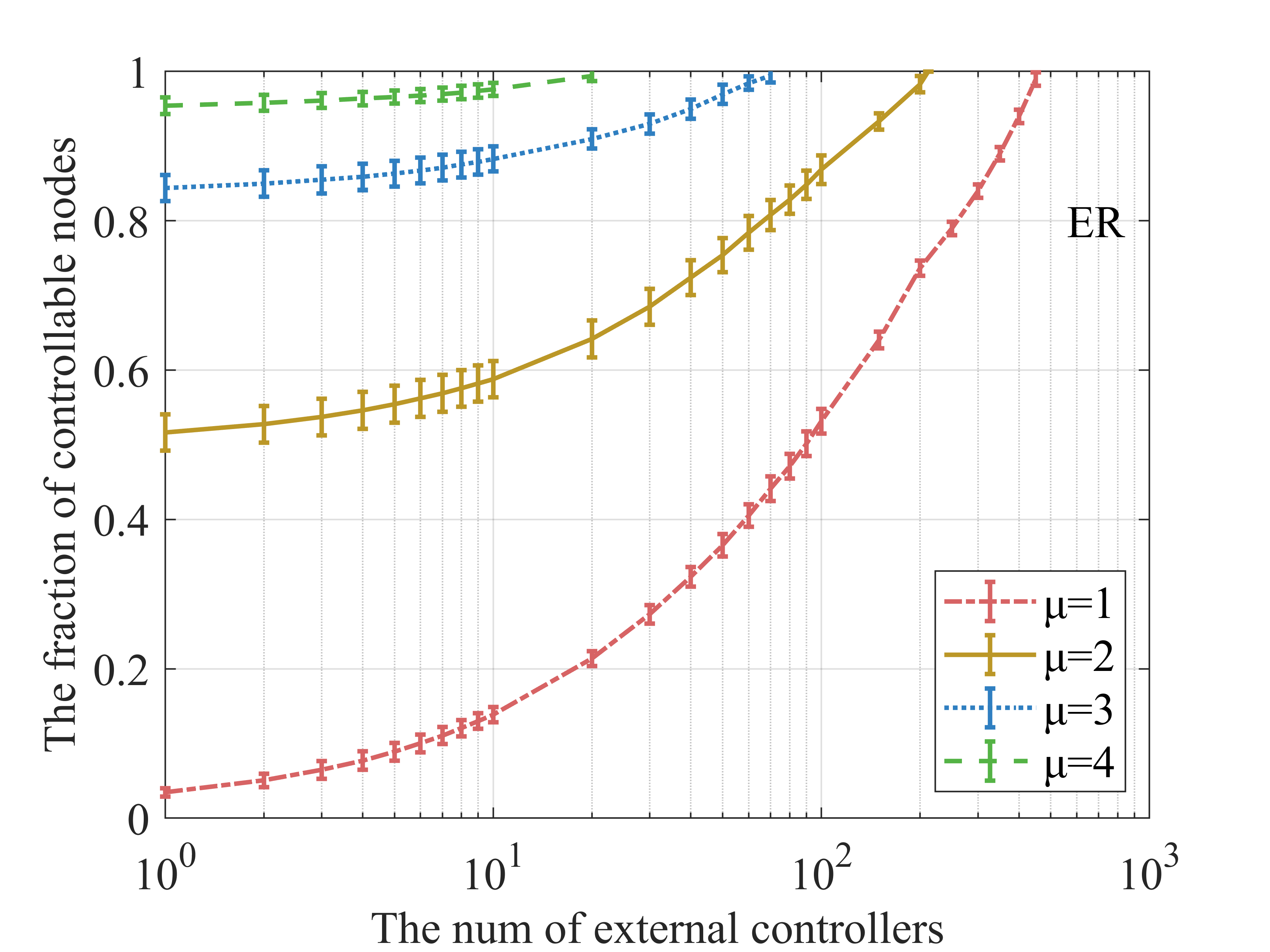}
\centering
\captionsetup{labelfont={color=black}}
	\caption[Minimum number of external control sources to guarantee the sufficient controllability in ER networks.]
	{The number of external controllers (log-scale) vs. the maximum fraction of network nodes being controllable in 1000-node ER networks. $\mu$ denotes the mean degree of the network.}
	\label{Fig.erctrl}
\end{minipage}
\hspace{.3cm}
\begin{minipage}{.48\textwidth}
    \centering
    \includegraphics[width=\textwidth]
{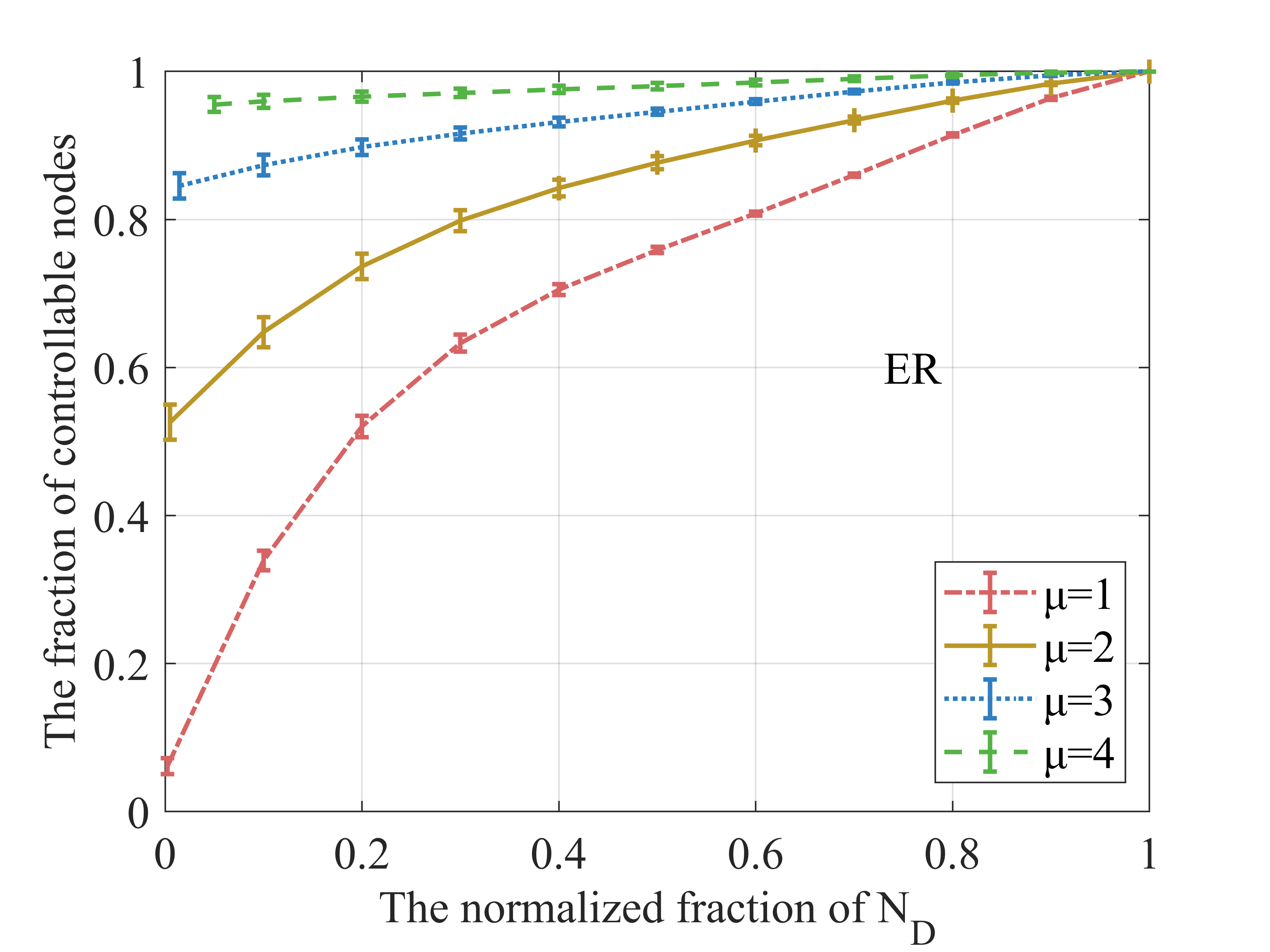}
	\captionsetup{labelfont={color=black}}
	\caption[The normalized fraction of driver nodes vs. the maximum fraction of network nodes being controllable in 1000-node ER networks. $\mu$ denotes the mean degree of the network.]
	{The normalized fraction of driver nodes vs. the maximum fraction of network nodes being controllable in 1000-node ER networks. $\mu$ denotes the mean degree of the network.}
	\label{Fig.ctrl_er_mus}
\end{minipage}

	% \includegraphics[width=0.45\textwidth]{Figures/final_ER_controlled_over_different_ctrlers2.png}
	% \captionsetup{labelfont={color=black}}
	% \caption[Minimum number of external control sources to guarantee the sufficient controllability in ER networks.]
	% {The number of external controllers (log-scale) vs. the maximum fraction of network nodes being controllable in 1000-node ER networks. $\mu$ denotes the mean degree of the network.}
	% \label{Fig.erctrl}
\end{figure}

We apply the MCFP algorithm in ER \cite{Erdoes1960}, BA \cite{BARABASI2003} as well as some real-life networks \cite{Chen2006, Milo2004, ulanowicz2012growth, doi:10.1080/15427951.2009.10129177, Colizza2007}.
The numbers of the external controllers versus the maximum fractions of network nodes being controllable in 1000-node ER random networks with the average nodal degree $\mu$ varying from 1 to 4 are shown in Fig. \ref{Fig.erctrl}. Simulations are conducted in 5 independently generated ER networks for each value of $\mu$, and error bars showing the standard deviation are plotted. It is observed that (i) when the number of external control sources is increased from 1 to $N_D$ (the minimum number of driver nodes needed to drive the whole network), the corresponding size of the maximum controllable nodes subset also increases; and (ii) the denser the network is, the same number of control sources can control the more nodes. In other words, dense networks are easier to be sufficiently controlled than sparse networks. This can be easily explained because longer paths or cycles that cover a larger number of nodes may be more easily identified in denser networks. When $\mu \geq 3$, a single controller may ensure the controllability of hundreds of nodes in a 1000-node ER network. When $\mu = 4$, a single controller can control almost the whole network, though the corresponding energy cost may be prohibitively high. 

We show the maximum fraction of network nodes being controllable as a function of the normalized fraction of driver nodes $N_\text{D}$ in 1000-node ER networks in Fig. \ref{Fig.ctrl_er_mus}. $N_\text{D}$ is the minimum number of controllers required for achieving the structural controllability of the whole network, which can be determined by the MM algorithm \cite{Liu2011}.
As can be observed, in denser networks, the maximum fraction of network nodes being controllable becomes much higher with the same normalized fraction of driver nodes; and the fraction of controllable nodes increases fast at the beginning and then slows down. These observations can be explained: at the beginning, a small number of extra-long paths or cycles can cover hundreds of nodes in a 1000-node ER network, especially in denser networks, while as the fraction of controllable nodes approaches 1, shorter paths or cycles may have to be used.

Note that in either sparse or relatively denser networks, the fractions of nodes being controllable are consistently much higher than the neutral expectation where controlling a fraction of network nodes requires the same fraction of the driver nodes needed for full control. Such an observation is different from that made in some studies on target controllability (e.g., \cite{Gao2014, Li2020}), where the performances are close to the neutral expectation. This can be explained as the target nodes in those studies were randomly selected while the controllable nodes selected by a sufficient control scheme generally do not have a random distribution; instead, these nodes tend to be connected, belonging to a certain number of control paths and cycles.

% \begin{figure}[H]
% 	\centering
% \includegraphics[width=0.45\textwidth]{Figures/final_ER_fraction_fraction.png}
% 	\captionsetup{labelfont={color=black}}
% 	\caption[The normalized fraction of driver nodes vs. the maximum fraction of network nodes being controllable in 1000-node ER networks. $\mu$ denotes the mean degree of the network.]
% 	{The normalized fraction of driver nodes vs. the maximum fraction of network nodes being controllable in 1000-node ER networks. $\mu$ denotes the mean degree of the network.}
% 	\label{Fig.ctrl_er_mus}
% \end{figure}

\begin{figure}[H]
	\centering
 \begin{minipage}{.48\textwidth}
	\includegraphics[width=\textwidth]{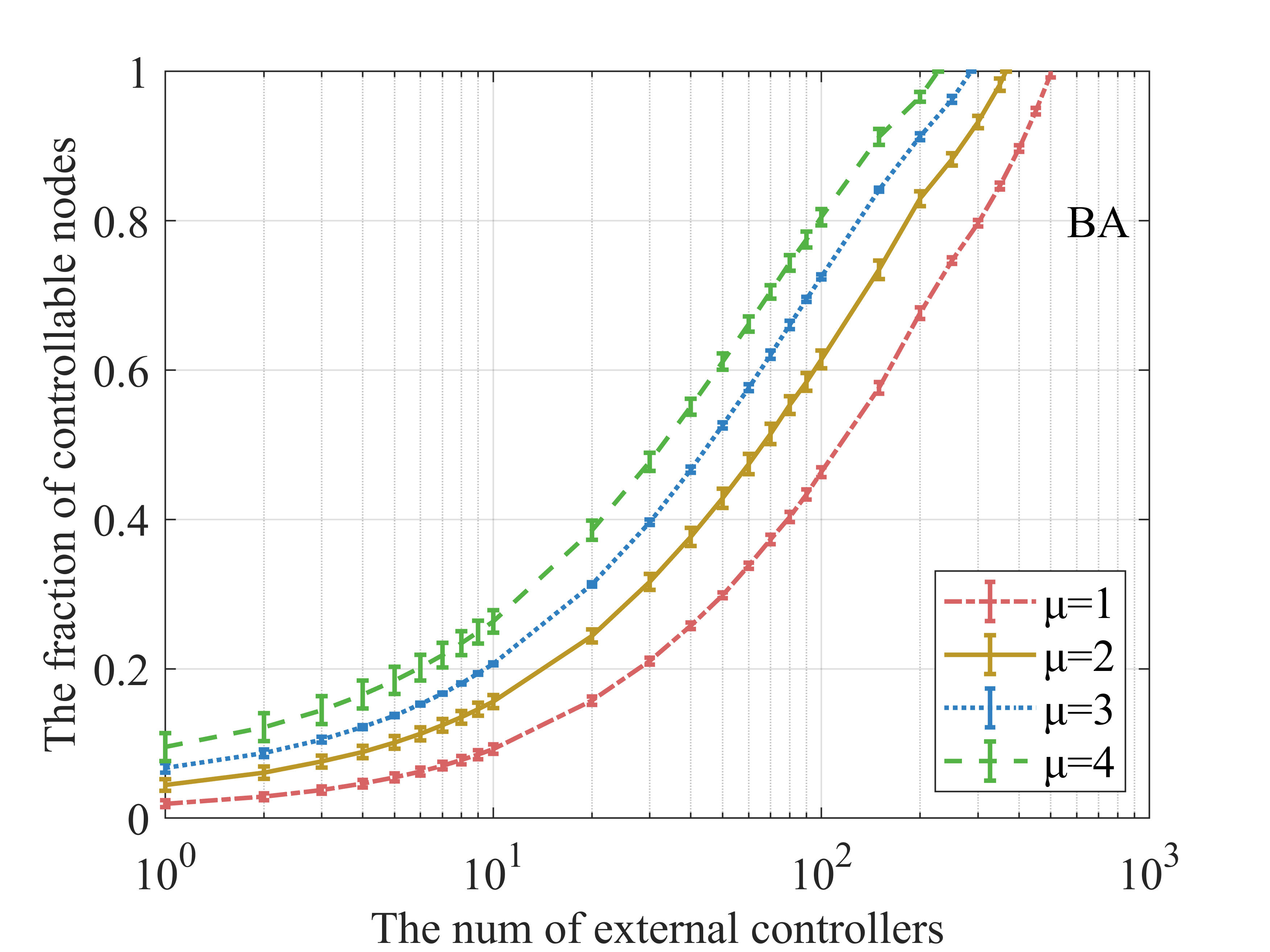}
	\captionsetup{labelfont={color=black}}
	\caption[Minimum number of external control sources to guarantee the sufficient controllability in BA networks.]
	{The number of external controllers (log-scale) vs. the maximum fraction of network nodes being controllable in 1000-node BA networks. $\mu$ denotes the mean degree of the network.}
	\label{Fig.bactrl}
 \end{minipage}
\hspace{.3cm}
\begin{minipage}{.48\textwidth}
    \centering
	\includegraphics[width = \textwidth]{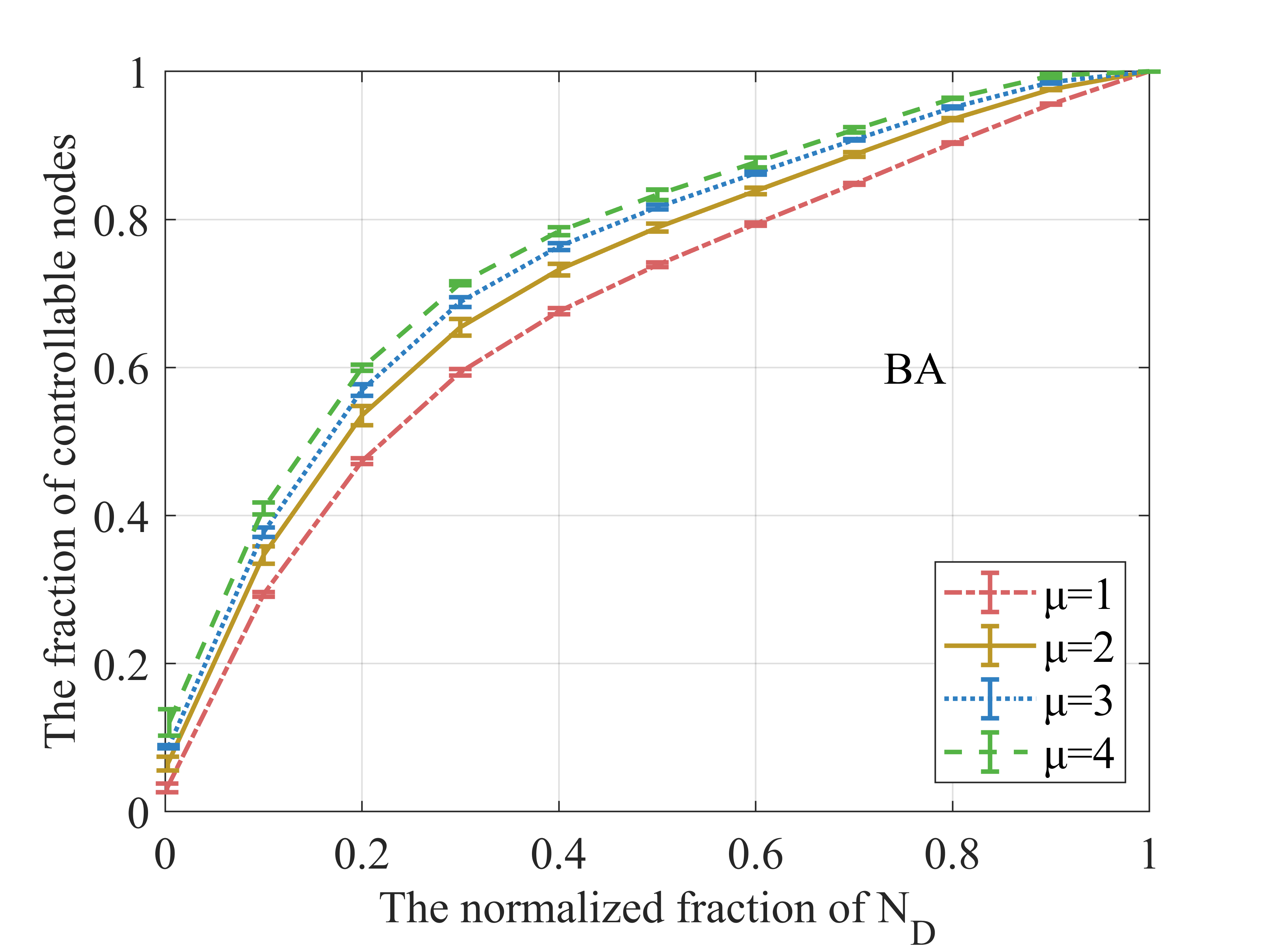}
	% \captionsetup{labelfont={color=blue}}
	\caption[The normalized fraction of driver nodes vs. the maximum fraction of network nodes being controllable in 1000-node BA networks. $\mu$ denotes the mean degree of the network.]
	{The normalized fraction of driver nodes vs. the maximum fraction of network nodes being controllable in 1000-node BA networks. $\mu$ denotes the mean degree of the network.}
	\label{Fig.ctrl_ba_mus}
\end{minipage}
\end{figure}

% \begin{figure}[H]
% 	\centering
% 	\includegraphics[width=3.3in]{Figures/final_BA_different_no_ctrl1.png}
% 	\captionsetup{labelfont={color=black}}
% 	\caption[Minimum number of external control sources to guarantee the sufficient controllability in BA networks.]
% 	{The number of external controllers (log-scale) vs. the maximum fraction of network nodes being controllable in 1000-node BA networks. $\mu$ denotes the mean degree of the network.}
% 	\label{Fig.bactrl}
% \end{figure}

% \begin{figure}[H]
% 	\centering
% 	\includegraphics[width=3.3in]{Figures/final_BA_frac_frac.png}
% 	% \captionsetup{labelfont={color=blue}}
% 	\caption[The normalized fraction of driver nodes vs. the maximum fraction of network nodes being controllable in 1000-node BA networks. $\mu$ denotes the mean degree of the network.]
% 	{The normalized fraction of driver nodes vs. the maximum fraction of network nodes being controllable in 1000-node BA networks. $\mu$ denotes the mean degree of the network.}
% 	\label{Fig.ctrl_ba_mus}
% \end{figure}

The numbers of external controllers versus the maximum fractions of network nodes being controllable in 1000-node BA random networks with $\mu$ varying from 1 to 4 are plotted in Fig. \ref{Fig.bactrl}. For each value of $\mu$, the MCFP algorithm is simulated in 5 independently generated BA networks. Compared to Fig. \ref{Fig.erctrl}, the main observation is that 
the numbers of external control sources requested to control the same fraction of nodes in BA networks are typically significantly higher than those in the corresponding ER networks with the same average nodal degree, except when $\mu = 1$. This is due to the different nodal degree distributions of the BA and ER networks. 
The nodal degree belongs to a Poisson distribution in ER network, whereas it has a power-law distribution in the BA network. Compared to an ER network, a BA network has a much larger portion of low-degree nodes in which long control paths or cycles may have lower chances of emerging. 

As shown in Fig. \ref{Fig.ctrl_ba_mus}, similar to that in the ER networks, in the BA random networks, the fraction of controllable nodes increases quickly with the number of controllers at the beginning; the increasing speed would then slow down. The fractions of controllable nodes are also consistently higher than the neutral expectation. Compared to Fig. \ref{Fig.ctrl_er_mus}, the differences caused by different connection densities, however, are much less significant than those in the ER networks. 

\begin{figure}[!t]
	\centering
 \begin{minipage}{.48\textwidth}
     \includegraphics[width=\textwidth]{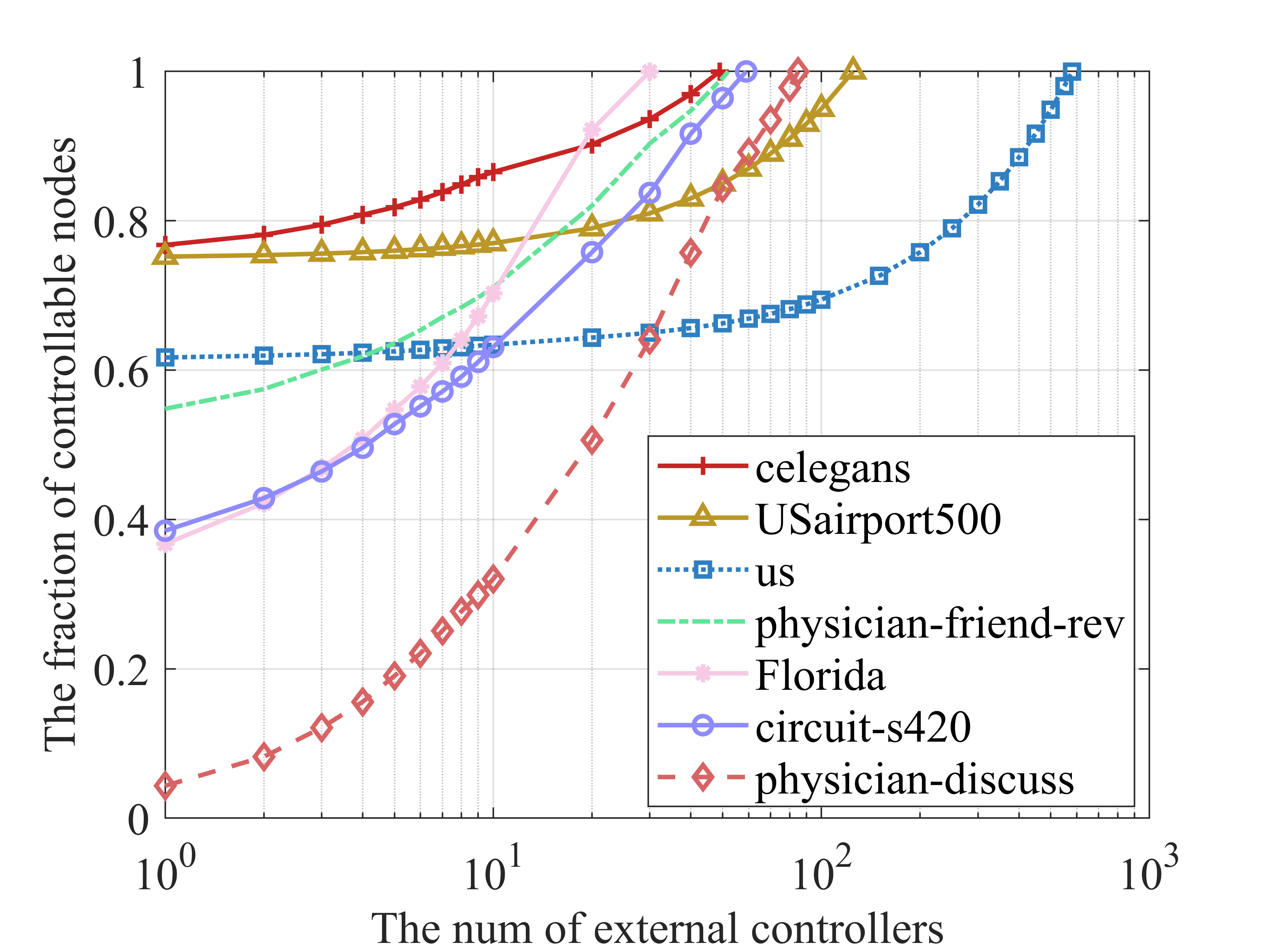}
	\captionsetup{labelfont={color=black}}
	\caption[Minimum number of external control sources to guarantee the sufficient controllability in real-life networks.]
	{The number of external controllers (log-scale) vs. the maximum fraction of network nodes being controllable in real-life networks.}
	\label{Fig.life_controllability}
 \end{minipage}
 \hspace{.3cm}
  \begin{minipage}{.48\textwidth}
     \centering
	\includegraphics[width=1.02\textwidth]{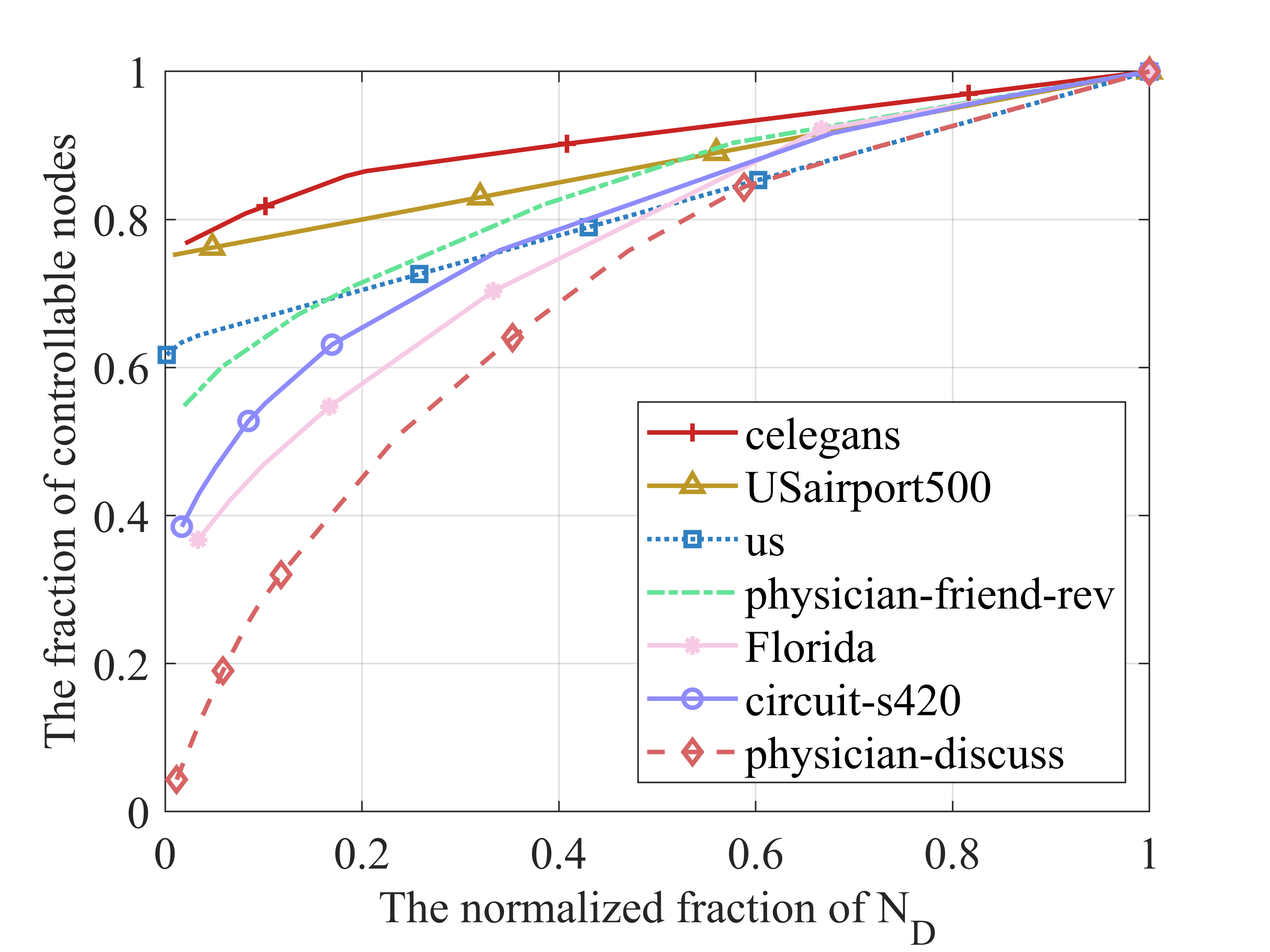}
	% \captionsetup{labelfont={color=blue}}
	\caption[The normalized fraction of driver nodes vs. the maximum fraction of network nodes being controllable in real-life networks.]
	{The normalized fraction of driver nodes vs. the maximum fraction of network nodes being controllable in real-life networks.}
	\label{Fig.life_normalized}
 \end{minipage}
	
\end{figure}

We also apply the MCFP algorithm on some real-life networks (C. elegans \cite{Chen2006}, USairport500 \cite{Colizza2007}, us \cite{Colizza2007}, physician-friend-review \cite{doi:10.1080/15427951.2009.10129177}, Florida \cite{ulanowicz2012growth}, circuit-s420 \cite{Milo2004}, and physician-discuss \cite{doi:10.1080/15427951.2009.10129177}). The numbers of external controllers versus the maximum fractions of network nodes being controllable are plotted in Fig. \ref{Fig.life_controllability}. It is observed that, basically, the denser a network is, the fewer controllers are required to control it, either fully or partially. 

The normalized fractions of driver nodes versus the maximum fractions of network nodes being controllable in real-life networks are plotted in Fig. \ref{Fig.life_normalized}. Similar to that in the ER/BA networks, in real-life networks, the fractions of controllable nodes increase rapidly at the beginning and then slow down, and the fractions of controllable nodes are consistently higher than the neutral expectation.

% \begin{figure}[!t]
% 	\centering
% 	\includegraphics[width=3.3in]{Figures/real_life_no4.png}
% 	\captionsetup{labelfont={color=black}}
% 	\caption[Minimum number of external control sources to guarantee the sufficient controllability in real-life networks.]
% 	{The number of external controllers (log-scale) vs. the maximum fraction of network nodes being controllable in real-life networks.}
% 	\label{Fig.life_controllability}
% \end{figure}

% \begin{figure}[!t]
% 	\centering
% 	\includegraphics[width=3.3in]{Figures/real_life_frac4.png}
% 	% \captionsetup{labelfont={color=blue}}
% 	\caption[The normalized fraction of driver nodes vs. the maximum fraction of network nodes being controllable in real-life networks.]
% 	{The normalized fraction of driver nodes vs. the maximum fraction of network nodes being controllable in real-life networks.}
% 	\label{Fig.life_normalized}
% \end{figure}

In the next section, we shall study the minimum-cost sufficient control problem which aims to drive a sufficient portion of the network nodes to any predefined state with a given number of control sources at the minimum cost.

\section{Minimum-cost Sufficient Control Problem}
\label{sec:Minimum-cost Sufficient Control Problem}
Considering only the sufficient controllability of complex networks is inadequate for sufficiently controlling real-world networks. An arguably even more critical issue may be the minimum-cost sufficient control problem, which drives a sufficient portion of the network to any predefined state with a given number of control sources at the minimum cost. It is known that the minimum-cost control problem is NP-hard \cite{Olshevsky2014, Tzoumas2016}. Hence it can be easily derived that the minimum-cost sufficient control problem is also NP-hard, as the former can be viewed as a special case of the latter one when $T=N$.

\subsection{Minimum-cost Sufficient Control Problem Formulation}
The minimum-cost sufficient control objective is to use a given number of controllers to drive the states of a sufficiently large portion of network nodes from $\mathbf{y}_{0} = \mathbf{y}(0) = C \mathbf{x}(0)$ to the origin $\mathbf{y}_{f} = \mathbf{y}(t_{f}) = \mathbf{0}$ during the time interval $[0, t_{f}]$, while minimizing the control cost $J = \mathbb{E}[\int_{0}^{t_{f}}\mathbf{u}^\text{T}(t)\mathbf{u}(t)dt]$.
Here $\mathbb{E}[\cdot]$ is an expectation function argument over all realizations of the random initial state $\mathbf{x}(0)$. Note that $\mathbf{u}(t)$, $B$ and $C$ are the decision variables to be determined by the optimization process, and their optimal solutions give the optimal input signals, the optimal selections of the driver nodes and the controlled nodes, respectively. 
In {\cite{Gao2018}}, Gao {\it et al.} have shown that, given that a system $(A, B, C)$ is output controllable, the minimum-cost control is achieved when
\begin{equation}
	\mathbf{u}(t) = -B^\text{T}e^{A^\text{T}(t_{f} - t)}C^\text{T} (CWC^\text{T})^{-1}Ce^{At_{f}}\mathbf{x}_{0}
\end{equation}
where $W = \int_{0}^{t_{f}}e^{At}BB^\text{T}e^{A^\text{T}t}dt$ is the controllability Gramian matrix. 

The minimum-cost sufficient control problem therefore can be formulated as:
\begin{equation}
	%\begin{small}
		\begin{aligned}
			&\underset{B, C}{\text{min}}
			&&J(B, C) = 
			e^{A t_{f}} e^{A^{\text{T}} t_{f}}\left(C^\text{T}\left[ CWC^\text{T} \right]^{-1} Ce^{At_{f}} \mathbb{E}(\mathbf{x}_{0} \mathbf{x}_{0}^\text{T}) e^{A^\text{T}t_{f}} \right) \\
			& \text{s.t.} && B \in \{0, 1\}^{N \times M},\\
   &&& \|B\|_{0} = M,\\
   &&& \operatorname{tr}\left(B^{\text{T}} B\right) = M, \\
   &&& C \in \{0, 1\}^{T \times N},\\
   &&& \|C\|_{0} = T,\\
   &&& \operatorname{tr}\left(C C^{\text{T}}\right)=T, \\
    &&&\operatorname{rank}(C) = T,\\
   &&&\operatorname{rank}[CB, CAB,\cdots, CA^{N-1}B] = T, \\
			% &&& \begin{array}{l}
				%  \quad \operatorname{tr}\left(B^{\text{T}} B\right) = M, \quad \|B\|_{0} = M, \quad \operatorname{rank}(B) = M,\\
				% , \quad \|C^{\text{T}}\|_{0} = T, \quad \operatorname{rank}(C^{\text{T}}) = T.
			% \end{array}
		\end{aligned}
		\label{E.optimalsufficientcontrol}
	%\end{small}
\end{equation}
where $\|\cdot\|_{0}$ denotes the $L_{\mathrm{0}}$ norm which counts the number of non-zero elements of a matrix, and $T$ is the number of nodes needed to be controllable. $M$ should be greater than the minimum number of controllers required by sufficient controllability. The full rank output controllability matrix implies that the output controllability Gramian matrix $CWC^{\text{T}}$ also be full rank \cite{Rugh1996}.
Since there is only one nonzero element in each row or column in $B$ or $C$, the selection of $M$ driver nodes and $T$ controllable nodes is hence clearly revealed.

Without loss of generality, it is assumed that each element of the initial state $\mathbf{x}_{0}$ is an identical independently distributed (i.i.d) variable with zero mean and variance one, i.e., $\mathbb{E}(\mathbf{x}_{0} \mathbf{x}_{0}^{\text{T}}) = I$.

\subsection{Extended $L_{\mathrm{0}}$-norm-constraint-based Projected Gradient Method}
We shall first introduce the ``extended $L_{\mathrm{0}}$-norm-constraint-based Projected Gradient Method" (eLPGM) algorithm as an extension of the LPGM algorithm \cite{Gao2018}. The main idea of the eLPGM algorithm is to analytically find the gradient of the control cost function, ${\frac{\partial J}{\partial B}}$ and ${\frac{\partial J}{\partial C}}$, and then iteratively project the negative gradient onto the manifold 
$\operatorname{tr}\left(B^{\text{T}}B\right)= M$, $\|B\|_{0}= M,$
and $\operatorname{tr}\left(C C^{\text{T}}\right)=T$, $\|C\|_{0}= T$ until the calculation results converge. The main difference between eLPGM and LPGM lies in the fact that eLPGM optimizes both $B$ and $C$ as variables, whereas LPGM only optimizes $B$. The following theorem presents the gradients of the control cost with respect to $B$ and $C$.
\begin{theorem}
\label{theorem_matrix_bc_derivative}
    The derivatives of the control cost with respect to $B$ and $C$ are
\begin{equation}
%	\begin{small}
		\begin{aligned}
			 \frac{\partial J(B, C)}{\partial B} = -2\int_{0}^{t_{f}} e^{A^{\text{T}} t} C^{\text{T}}\left(C W C^{\text{T}}\right)^{-\text{T}} C e^{A t_{f}} e^{A^{\text{T}} t_{f}}C^{\text{T}}\left(C W C^{\text{T}}\right)^{-\text{T}} C e^{A t} B d t
		\end{aligned}
	%\end{small}
	\label{E.gradient_B}
\end{equation}
and
\begin{equation}
%	\begin{small}
		\begin{aligned}
			 \frac{\partial J(B, C)}{\partial C} = 
   -2\left(C W C^{\text{T}}\right)^{-1}C e^{A t_{f}} e^{A^{\text{T}} t_{f}}C^{\text{T}}\left(C W C^{\text{T}}\right)^{-1}CW^{\text{T}} + 2\left(C W C^{\text{T}}\right)^{-1}Ce^{A t_{f}} e^{A^{\text{T}} t_{f}}.
   % -2 W C^{\text{T}}\left(C W C^{\text{T}}\right)^{-1} C e^{A t_{f}} e^{A^{\text{T}} t_{f}} C^{\text{T}}\left(C W C^{\text{T}}\right)^{-1}
			% +2 e^{A t_{f}} e^{A^{\text{T}} t_{f}} C^{\text{T}}\left(C W C^{\text{T}}\right)^{-1}.
		\end{aligned}
	%\end{small}
	\label{E.gradient_C}
\end{equation}
    
\end{theorem}
\begin{proof}
    Please refer to \ref{proof_matrix_bc_derivative}.
\end{proof}

% By varying (\ref{E.optimalsufficientcontrol}) with respect to $B$ and $C$ respectively, we find that 

% By applying (\ref{E.optimalsufficientcontrol})-(\ref{E.gradient_C}) to the eLPGM algorithm, we solve for the energy-optimal input matrix and output matrix. 

Before discussing the details of the eLPGM algorithm, we firstly define the projection operator $\mathcal{F}(H, m_0)$ which projects a given matrix $H = (h_{ij}) \in \mathbb{R}^{N \times m_0} (m_0 \leq N)$ onto the manifold $H \in \{0,1\}^{N \times m_0}, \operatorname{tr}\left(H^\text{T} H\right) = m_0$, $\|H\|_{0} = m_0$, $\operatorname{rank}(H) = m_0$. The output of the projection operator is denoted by $H^{L_{0}}$. Such an operator is needed for carrying out iterative projected gradient calculations for solving the optimization problem (\ref{E.optimalsufficientcontrol}). As that will be shown in detail in Algorithm \ref{alg:eLPGM Algorithm}, the projection operator is applied on both matrix $B$ and matrix $C$. The function of the projection operator is to select $m_0$ nodes, and then construct the corresponding $H^{L_{0}}$ based on the selected nodes. 

The node selection process involves computing the {\it importance index} $h_i$ for node $v_i$ by summing the absolute values of the entries in the $i^{\text{th}}$ row of the matrix $H$, i.e.,
\begin{equation}
	%\begin{small}
		h_{i}= \sum_{j}\left|h_{i j}\right|, \quad (i=1, \ldots, N).
	%\end{small}
\end{equation}
Selecting nodes with relatively higher importance index values as driver nodes or controlled nodes tends to lower the sufficient control cost. This is because in the calculation results of each iteration, either $B$ or $C$ is of real values, while the projection operator maps the matrix into a Boolean matrix reflecting the node selections. Hence it makes sense to give a node with a higher importance index value a higher chance of being assigned a value 1, and vice versa. To introduce some necessary randomness, we pick the first $m_0 + m_1$ ($m_1 \geq 1$) nodes with the largest importance index values for further selections. Existing results (\cite{Gao2018}) and our own experiences show that introducing a moderate level of randomness by adopting a small value $m_1$ helps improve the algorithm performance. The candidate node set consisting of these $m_0 + m_1$ nodes is denoted as $V_{\text{cand}}$. We normalize the importance index of each node in $V_{\text{cand}}$ as
\begin{equation}
		\theta_{i} = \frac{h_{i}}{\sum_{v_j\in V_{\text{cand}}} h_{j} } \text{,} \quad v_{i} \in V_{\text{cand}}.
\end{equation}
After randomly selecting a node from $V_{\text{cand}}$ based on its probability $\theta_{i}$, we add the node to a set of selected nodes denoted as $V_{\text{sel}}$ and remove it from the candidate node set $V_{\text{cand}}$. We then normalize the probability of the remaining nodes in $V_{\text{cand}}$. The procedure described above is repeated iteratively until $m_0$ nodes have been selected. Finally, the resulting matrix $H^{L_{0}} = (h^{L_{0}}_{ij}) \in \{0,1\}^{N \times m_0}$ is constructed by setting $h^{L_{0}}_{ij}=1$ if node $v_i$ is the $j^{\text{th}}$ element in $V_{\text{sel}}$, and zero otherwise. The pseudocode of the projection operator $\mathcal{F}(H, m_0)$ is illustrated in lines 8 - 18 in Algorithm \ref{alg:eLPGM Algorithm}.

\begin{algorithm}[H]
	\begin{small}
		\caption{eLPGM Algorithm}
		\begin{algorithmic}[1]
			\renewcommand{\algorithmicrequire}{\textbf{Input:}}
			\renewcommand{\algorithmicensure}{\textbf{Output:}}
			\Require $A$, $N$, $M$, $T$, the maximum iteration number $k_{\text{max}}$
			\Ensure $B^*$, $C^*$
			\State Initialize $B_{k = 0}^{L_0}$, and $C_{k = 0}^{L_0}$ such that $\left(A, B_{0}^{L_0},C_{0}^{L_0}\right)$ is output controllable;
			and $B^* = B_{0}^{L_{0}}$, $C^* = C_{0}^{L_{0}}$ and $J^* =J\left(B^*,C^*\right)$.
			\For{$k = 0,1,\dots, k_{\text {max}}$}
			\State Update $B_{k+1} = B_{k}^{L_{0}}-\eta_{k} \frac{\partial J(B_k^{L_{0}}, C_k^{L_{0}})}{\partial B_k^{L_{0}}}$,
			\Statex \qquad \qquad \quad $B_{k+1}^{L_{0}} = \mathcal{F}\left(B_{k+1}, M\right)$,
			\Statex \qquad \qquad \quad $C_{k+1} = C_{k}^{L_{0}}-\alpha_{k} \frac{\partial J(B_{k+1}^{L_{0}}, C_k^{L_{0}})}{\partial C_k^{L_{0}}}$,
			\Statex \qquad \qquad \quad $C_{k+1}^{L_{0}} = (\mathcal{F}(C_{k+1}^\text{T}, T))^\text{T}$.
			\If{$J\left(B_{k+1}^{L_{0}}, C_{k+1}^{L_{0}}\right)<J^*,$}
			\State $B^* = B_{k+1}^{L_{0}}$,\quad $C^* = C_{k+1}^{L_{0}}$,
			\Statex \qquad \quad $J^* = J\left(B_{k+1}^{L_{0}}, C_{k+1}^{L_{0}} \right)$.
			\EndIf
			\EndFor
			
			% \\\Return $B_{\text {best}}$, $C^T_{\text{best}}$
			\Statex
			\Function{$\mathcal{F}$}{$H, m_0$}
			\State Obtain the nodal importance index $h_{i}= \sum_{j}\left|h_{i j}\right|, \quad (i=1, \ldots, N)$. 
			\State Initialize $V_{\text{cand}}$ consisting of $m_0+m_1$ nodes with the largest importance index values ($1 \leq m_1 $); $V_{\text{sel}} = \emptyset$.
			\While{$|V_{\text{sel}}| < m_0$} 
			\State Normalize the importance index  
  $\theta_{i} = \frac{h_{i}}{\sum_{V_j\in V_{\text{cand}}} h_{j} } \text{,} \quad v_{i} \in V_{\text{cand}}$
			\State Randomly select a node $v_k$ from $V_{\text {cand}}$ based on the probability proportional to $\theta_{k}$
			\State Update $V_{\text {sel}} = V_{\text {sel}} \cup v_k$,\quad $V_{\text {cand}} = V_{\text {cand}} \setminus v_k$
			\EndWhile
			\State $H^{L_{0}}_{ij}=1$, if $i$ is the node index of the $j^{\text{th}}$ selected node; $H^{L_{0}}_{ij}=0$, otherwise. ($1 \leq j \leq m_0$)
			\\\Return $H^{L_{0}}$
			\EndFunction
		\end{algorithmic}
		\label{alg:eLPGM Algorithm} 
	\end{small}
\end{algorithm}

The main steps of the eLPGM algorithm are introduced as follows. As shown in Algorithm \ref{alg:eLPGM Algorithm}, line 1 initializes $B_{k=0}^{L_0}$ and $C_{k=0}^{L_0}$ as binary matrices satisfying certain conditions to ensure output controllability. We denote $B_k$ by $B_k^{L_0}$ if it satisfies $B_k \in \{0, 1\}^{N \times M}$, $\operatorname{rank}(B_k) = M$, $\operatorname{tr}(B_k^{\text{T}} B_k) = M$, and $|B_k|_0 = M$. Similarly, we denote $C_k$ by $C_k^{L_0}$ if it satisfies $C_k \in \{0, 1\}^{T \times N}$, $\operatorname{rank}(C_k) = T$, $\operatorname{tr}(C_k C_k^{\text{T}}) = T$, and $|C_k|_0 = T$.
Lines 2 - 7 perform a for loop to iteratively update the optimal input matrix $B^*$ and output matrix $C^*$. In the $k^{\text{th}}$ iteration, 
the projected gradient descent method is applied to update $B_{k+1}^{L_{0}}$ and $C_{k+1}^{L_{0}}$ as follows,
\begin{equation}
	%\begin{small}
		\begin{aligned}
			&B_{k+1} = B_{k}^{L_{0}}-\eta_{k} \frac{\partial J(B_k^{L_{0}}, C_k^{L_{0}})}{\partial B_k^{L_{0}}},\\
			&B_{k+1}^{L_{0}} = \mathcal{F}\left(B_{k+1}, M\right),\\
			&C_{k+1} = {C_{k}}^{L_{0}}-\alpha_{k} \frac{\partial J(B_{k +1}^{L_{0}}, C_k^{L_{0}})}{\partial C_k^{L_{0}}},\\
			&{C_{k+1}^{L_{0}}} = (\mathcal{F}(C_{k+1}^\text{T}, T))^\text{T}.
		\end{aligned}
	%\end{small}
\end{equation}
where $\eta_k$ and $\alpha_k$ denote the learning rates for optimizing $B$ and $C$, respectively. $\eta_k$ and $\alpha_k$ are searched by the golden section method \cite{Press2007}.
The gradients are calculated according to (\ref{E.gradient_B}) and (\ref{E.gradient_C}), respectively. 
Then lines 4 - 6 are for finding the local minima of the control cost. Remarkably, the eLPGM algorithm is guaranteed to converge since the optimal control cost $J^*$ obtained in each iteration monotonically non-increases, as shown in line 4.
Finally, we obtain the solutions $B^*$ and $C^*$ corresponding to the local minima of the control cost.

\subsection{Evenly Divided Control Paths Algorithm}
Due to its high computational complexity, the eLPGM algorithm cannot be applied to large-scale networks, just like other projected gradient descent algorithms \cite{Li2015, Gao2018, Li2018, Ding2017}. We propose a low complexity heuristic graph algorithm termed ``Evenly Divided Control Paths" (EDCP) algorithm to tackle large-scale problems. 

It has been observed that the control cost is dominated by the longest control path \cite{Wang2017} and the optimal driver nodes tend to divide elementary topologies into largely equal lengths \cite{Li2015, Li2018, Ding2017}. Table \ref{table.cost} demonstrates the control costs when one controller is used to control a unidirectional, one-dimensional string network consisting of a varying number of nodes. It is observed that control cost scales exponentially with the control path length, which is consistent with the previous observations in \cite{Wang2017, Li2015, Ding2017}. 
\begin{table}[!t]
	%\renewcommand{\arraystretch}{1.6}
	\caption{Control cost with different control path lengths}
	\label{table.cost}
	\centering
	% Some packages, such as MDW tools, offer better commands for making tables
	% than the plain LaTeX2e tabular which is used here.
	\resizebox{0.8\textwidth}{!}{
		\begin{tabular}{lcccccccccc}
			\hline	Length: & 1 & 2 & 3 & 4 & 5 & 6 & 7 & 8 & 9 & 10\\
			\hline 	Cost: & 0.5 & 4 & 51 & 1.76E3 & 1.11E5 & 1.10E7 & 1.57E9 & 3.07E11 & 7.84E13 & 2.54E16\\
			\hline 
	\end{tabular}}
\end{table}
Motivated by such observations, we propose a simple graph-based EDCP algorithm, of which the main idea is to set control paths to be of nearly equal length as much as possible while ensuring sufficient control of the network. 

Before discussing the details of the EDCP algorithm, we first define a function $\mathscr{E}(q, d)$ to calculate the associate cost of using $d$ controllers to control a unidirectional, one-dimensional network consisting of $q$ nodes. A one-dimensional network consisting of $q$ nodes can be represented as $D(V, E)$ where $V = \{v_1, v_2, ..., v_q\}$ is the set of nodes and $E = \{(v_i, v_{i+1}): i = 1, 2, \cdots, q-1\}$ is the set of edges connecting adjacent nodes. The main steps of $\mathscr{E}(q, d)$ are as follows:

(i) The $q$-node path is evenly divided into $d$ parts such that their lengths are
\begin{equation}
	%\begin{small}
		\underbrace{\left\lfloor\frac{q}{d}\right\rfloor + 1,\left\lfloor\frac{q}{d}\right\rfloor + 1,\cdots\left\lfloor\frac{q}{d}\right\rfloor + 1}_{q \bmod d}, \underbrace{ \left\lfloor\frac{q}{d}\right\rfloor, \left\lfloor\frac{q}{d}\right\rfloor,\cdots,\left\lfloor\frac{q}{d}\right\rfloor}_{d - (q \bmod d )} 
	%\end{small}
\end{equation}
respectively, where $\lfloor \cdot \rfloor$ denotes the floor function.

(ii) The corresponding costs for each part are found in Table \ref{table.cost} based on their lengths, and the costs are summed. 
The function $\mathscr{E}(q, d)$ returns the summation of the costs.

The EDCP algorithm starts with a number of control paths and control cycles identified by the MCFP algorithm. To avoid confusion in later discussions, we term the control paths as paths, and the control cycles as cycles. And the path is divided into sub-paths by driver nodes, with each sub-path having a driver node as its initial node. The initialization step of the EDCP algorithm is to merge cycles into paths, where some or all paths later can be divided into the requested number of sub-paths for fulfilling sufficient control. Specifically, we break the longest cycle $\mathcal{c}_{\text{long}} \ in {\mathcal{C}}$ by removing one of its links. The broken cycle is denoted as $\mathcal{c}'_{\text{long}}$ and we connect it to the shortest path $\mathcal{p}_{\text{sht}} \in \mathcal{P}$, i.e., $\mathcal{p}_{\text{sht}} = \mathcal{p}_{\text{sht}} \cup \mathcal{c}'_{\text{long}}$. This process is repeated until there is no cycle left. In this way we shall obtain the path set $\mathcal{P}$. For each path $\mathcal{p_i} \in \mathcal{P}$, we initialize its undivided part as itself, i.e., $l_i = \mathcal{p_i}$.

To evenly divide $T$ into $M$ integers, we obtain a set 
\begin{equation}
	%\begin{small}
		Q = \Biggl\{\underbrace{\left\lfloor\frac{T}{M}\right\rfloor + 1,\left\lfloor\frac{T}{M}\right\rfloor + 1,\cdots\left\lfloor\frac{T}{M}\right\rfloor + 1}_{T \bmod M}, \underbrace{ \left\lfloor\frac{T}{M}\right\rfloor, \left\lfloor\frac{T}{M}\right\rfloor,\cdots,\left\lfloor\frac{T}{M}\right\rfloor}_{M - (T \bmod M )} \Biggl\}.
	%\end{small}
\end{equation}
We shall now start to assign the driver nodes along the paths in order to create sub-paths of roughly equal length until at least $T$ nodes are controllable. Specifically, for each element $q_i$ in $Q$, where $m = 1,2,\cdots,M$, we find the path $\mathcal{p}_j \in \mathcal{P}$ with the longest undivided part $l_j$. We set the first node in $l_l$ as a driver node. If $q_i \leq |\mathcal{p}_j|$, we update the undivided part $l_j$ by removing its first $q_m$ nodes. In this way, we get a sub-path with a length of $q_i$.
If $q_m > |\mathcal{p}_j|$, we set the first node in $l_{j}$ as a driver node and update $l_j = \emptyset$. In this way, we get a sub-path with a length of $|\mathcal{p}_j|$. After traversing $Q$, the number of controllable nodes may be lower than $T$ if certain sub-paths shorter than $q_m$ have been selected. For such a case, we shall repetitively select the path with the longest undivided part and set its first node as a drive node until at least $T$ nodes are controllable. 
Note that the number of controllable nodes may exceed $T$ when the last control path is selected. The method for releasing the extra controllable nodes when such is needed shall be discussed later.
\begin{algorithm}[H]
	\begin{small}
		\caption{EDCP Algorithm}
		\begin{algorithmic}[1]
			\renewcommand{\algorithmicrequire}{\textbf{Input:}}
	     \renewcommand{\algorithmicensure}{\textbf{Output:}}
			\Require Path set $\mathcal{P}$, cycle set $\mathcal{C}$ obtained by MCFP algorithm, $M$ and $T$
			%		control cost with different control path lengths
			\Ensure $B$ and $C$
			\While{${\mathcal{C}} \neq \emptyset$}
			\State Break the longest cycle $\mathcal{c}_\text{long} \in {\mathcal{C}}$, denote the broken cycle as $\mathcal{c}'_\text{long}$, update $\mathcal{C} = \mathcal{C} \setminus \mathcal{c}_\text{long}$ 
  \State Connect $\mathcal{c}'_\text{long}$ to the shortest path $\mathcal{p}_\text{sht} \in \mathcal{P}$, update $\mathcal{p}_\text{sht} = \mathcal{c}_\text{long} \cup \mathcal{p}_\text{sht}$
			\EndWhile	
			\State Initialize $n_{\text{c}} = 0$; and for each path $\mathcal{p}_k \in \mathcal{P}$, initialize the number of driver nodes $d_k = 0$, the undivided part $l_k = p_k$
			\State Obtain the set $Q = \{q_1, \cdots, q_M \} = \biggl\{\underbrace{\left\lfloor\frac{T}{M}\right\rfloor + 1,\cdots\left\lfloor\frac{T}{M}\right\rfloor + 1}_{T \bmod M}, \underbrace{ \left\lfloor\frac{T}{M}\right\rfloor, \cdots,\left\lfloor\frac{T}{M}\right\rfloor}_{M - (T \bmod M )} \biggl\}$ which divides $T$ into $M$ integers
			\For{$q_i \in Q$}
			\State Find the path $\mathcal{p}_{j} \in \mathcal{P}$ with the longest $l_{j}$, and set the first node in $l_{j}$ as a driver node and update $d_j = d_j +1$
               % \State 
			\If {$q_i \leq |{\mathcal{p}_{j}}|$}
			\State Update $l_j = l_j \setminus $ \{first $q_m$ nodes\}, $n_{\text{c}} = n_{\text{c}} + q_i$
			%		\State Update the length of $p_l$ as $|p_l| \leftarrow |p_l| - dp$
			\Else 
			\State Update $l_j = \emptyset$, $n_{\text{c}} = n_{\text{c}} + |\mathcal{p}_j|$
			%		\State Update the length of $p_l$ as $|p_l| \leftarrow 0$
			\EndIf
			%		\State Keep track of the path index where the $rod$ is from 
			\EndFor
			\While {$n_{\text{c}} < T$}
			\State Find the path $\mathcal{p}_{i} \in \mathcal{P}$ with the longest $l_{i}$, and set the first node of ${l}_{i}$ as a driver node
			\State Update $d_i = d_i +1$, $l_i = \emptyset$, $n_{\text{c}} = n_{\text{c}} + |l_i|$
			%		\State Update the length of $p_l$ as $|p_l| \leftarrow 0$ 
			\EndWhile
			%		\While {$\sum_{k= 1}^{|\mathcal{P}|} d_k > M$}
			\While {$\sum_l d_l > M$}
			\For{$\mathcal{p}_k$ in $\mathcal{P}$}
			\State Calculate the increased associate control cost $E'_k = \mathscr{E}(|\mathcal{p}_k \setminus l_k|, d_k) - \mathscr{E}(|\mathcal{p}_k \setminus l_k|, d_k - 1)$
			\EndFor 
			\State Release the driver node from the path with the least increase in control cost. Denote this path as $\mathcal{p}_m$. Re-assign the $d_m-1$ controllers along this path and update $d_m = d_m - 1$
			%		Obtain the stem $\mathcal{l}^*_{k = m}$ of which the increased cost $E'$ is the smallest in $\mathbf{\mathcal{P}^*}$
			%		\State Update $d_{m} \leftarrow d_{m}-1$, and then reassign the $d_{m}$ driver nodes in $\mathcal{l}^*_{m}$ as evenly as possible 
			\EndWhile
			\While {$n_{\text{c}} > T$}
			\State Release one node from the longest sub-path, and update $n_{\text{c}} = n_{\text{c}} -1$
			%		\State Delete the start node in $\mathcal{l}^*_{n}$ and reassign the $d_{n}$ driver nodes in $\mathcal{l}^*_{n}$ as evenly as possible 
		
			\EndWhile
			%		\State $B$ and $C$ are derived from control paths
			% \\
			% \Return $B$ and $C$ 
			\Statex
			% \Statex Function of calculating the control cost 
			% \Function{$\mathscr{E}$}{$q, d$}
			% \State Evenly divide the $q$-node path into $d$ parts. $\mathcal{S}$ represents the set of these parts' lengths, denoted as
			% $\mathcal{S} = \biggl\{\underbrace{\left\lfloor\frac{q}{d}\right\rfloor + 1,\cdots\left\lfloor\frac{q}{d}\right\rfloor + 1}_{q \bmod d}, \underbrace{ \left\lfloor\frac{q}{d}\right\rfloor, \left\lfloor\frac{q}{d}\right\rfloor,\cdots,\left\lfloor\frac{q}{d}\right\rfloor}_{d - (q \bmod d )} \biggl\}$. 
   % $\mathcal{E}$ represents the set of these parts' associate costs according to Table \ref{table.cost}, denoted as $\mathcal{E} = \{\mathcal{e}_1, \cdots, \mathcal{e}_d\}$
			% % \For{each sub-part length $s_l$ ($l = 1,2,\cdots,d$) in $\mathbf{S}$}
	
			% % \EndFor
			% \State Calculate the cost $E = \sum_{l = 1}^{d} \mathcal{e}_l$.
			% \\\Return $E$
			% \EndFunction
			% 		\algstore{myalg}
		\end{algorithmic} 
		\label{alg:EDCP Algorithm} 
	\end{small}
\end{algorithm}

\begin{figure}[t!]
	\centering
	\begin{subfigure}[b]{0.3\textwidth}
		\includegraphics[width= \textwidth]{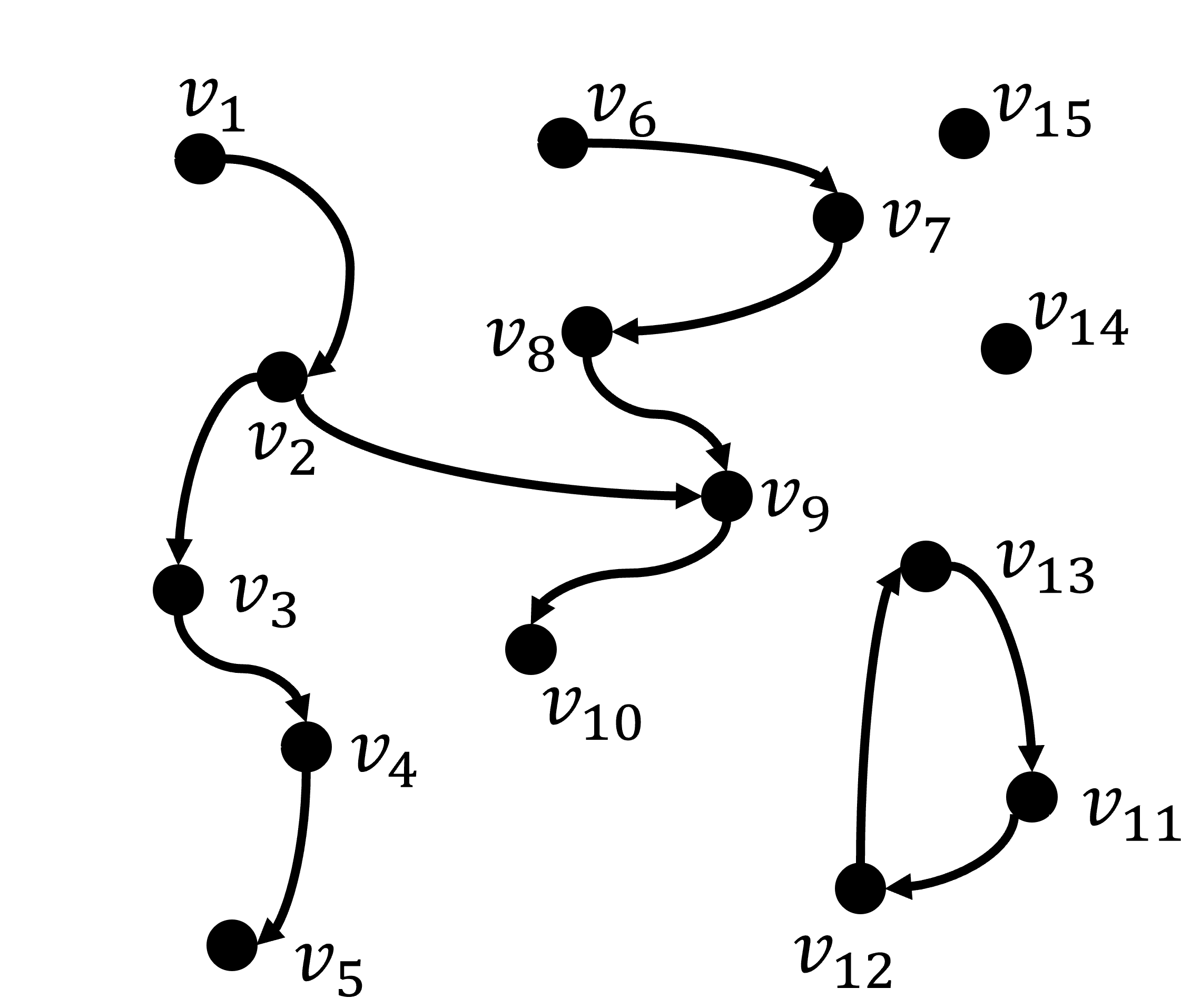}
		\caption{}
		\label{Fig.sub.edcp1}
	\end{subfigure}
 \hspace{2cm}
	%	\subcaptTionbox{a}{\includegraphics[width=0.20\textwidth]{Figures/1_a.png}\label{Fig.sub.edcp1}}%
	\begin{subfigure}[b]{0.3\linewidth}
		\includegraphics[width=1\linewidth]{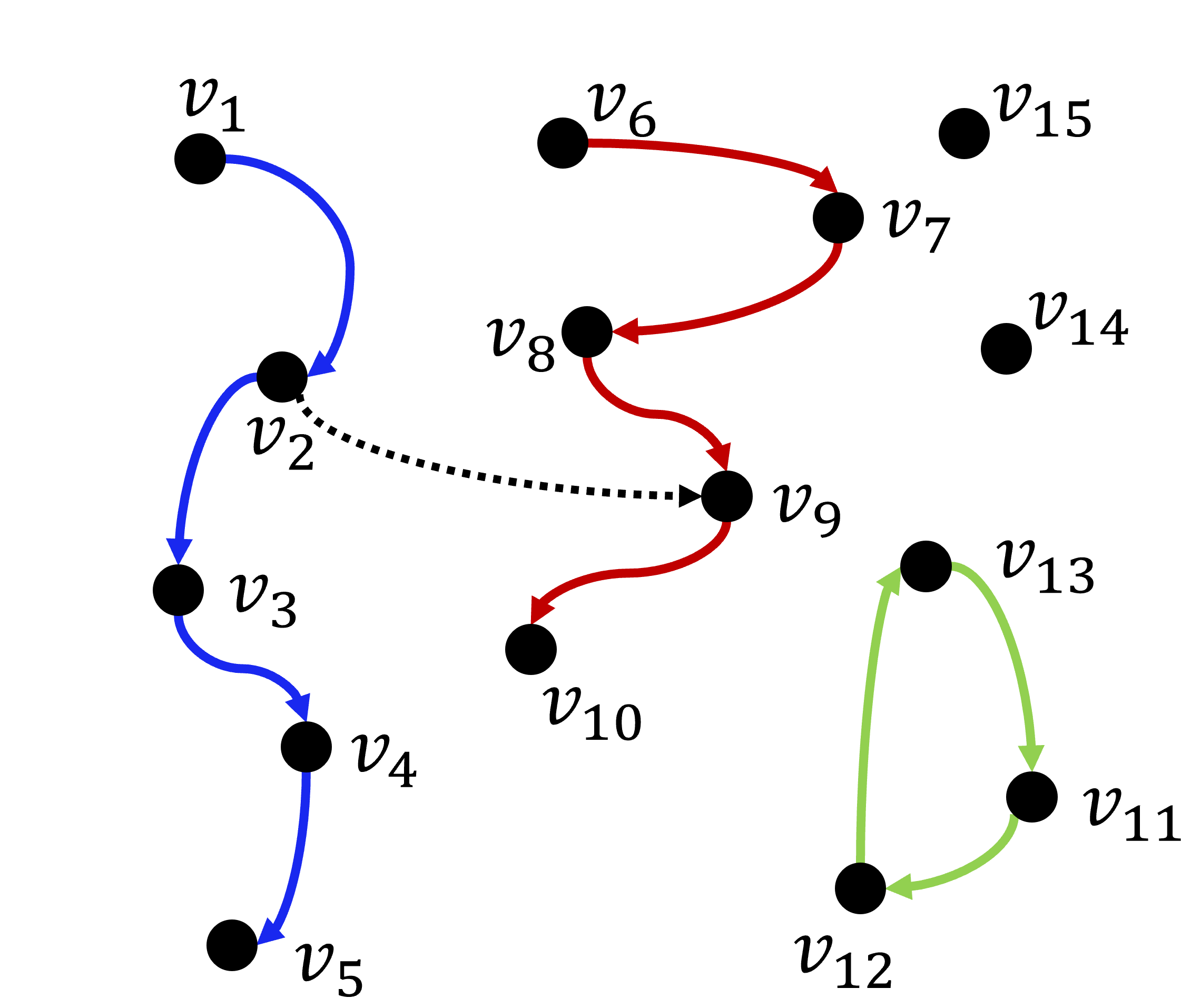}
		\caption{\label{Fig.sub.edcp2}}
	\end{subfigure}
 \hspace{0cm} \\
	\begin{subfigure}[t]{0.3\linewidth}
		\includegraphics[width=1\linewidth]{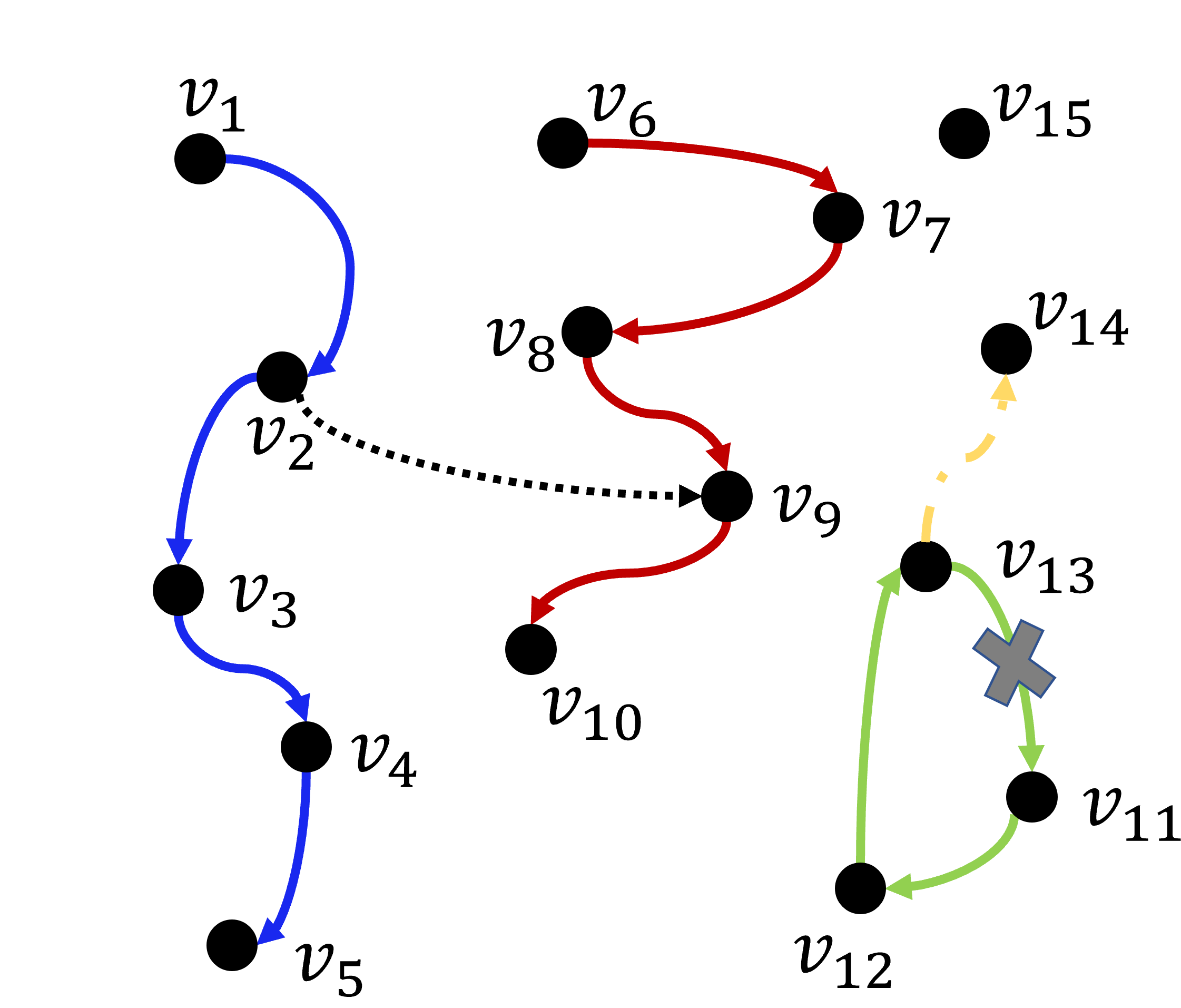}
		\caption{\label{Fig.sub.edcp3}}
	\end{subfigure}
 \hspace{2cm}
	\begin{subfigure}[t]{0.3\linewidth}
		\includegraphics[width=1\linewidth]{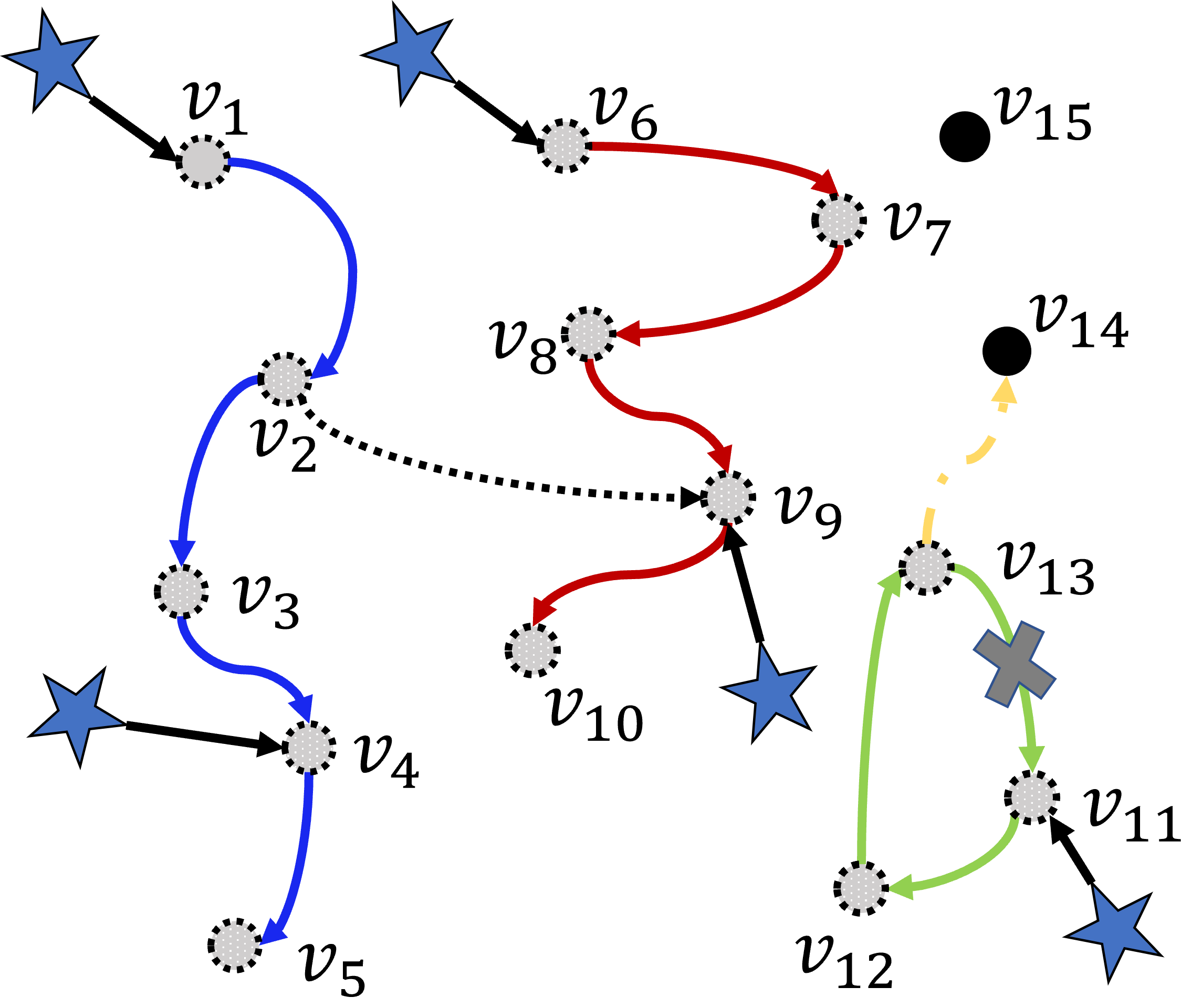}
		\caption{\label{Fig.sub.edcp4}}
	\end{subfigure}
 \hspace{0cm} \\
	\begin{subfigure}[t]{0.3\linewidth}
		\includegraphics[width=1\linewidth]{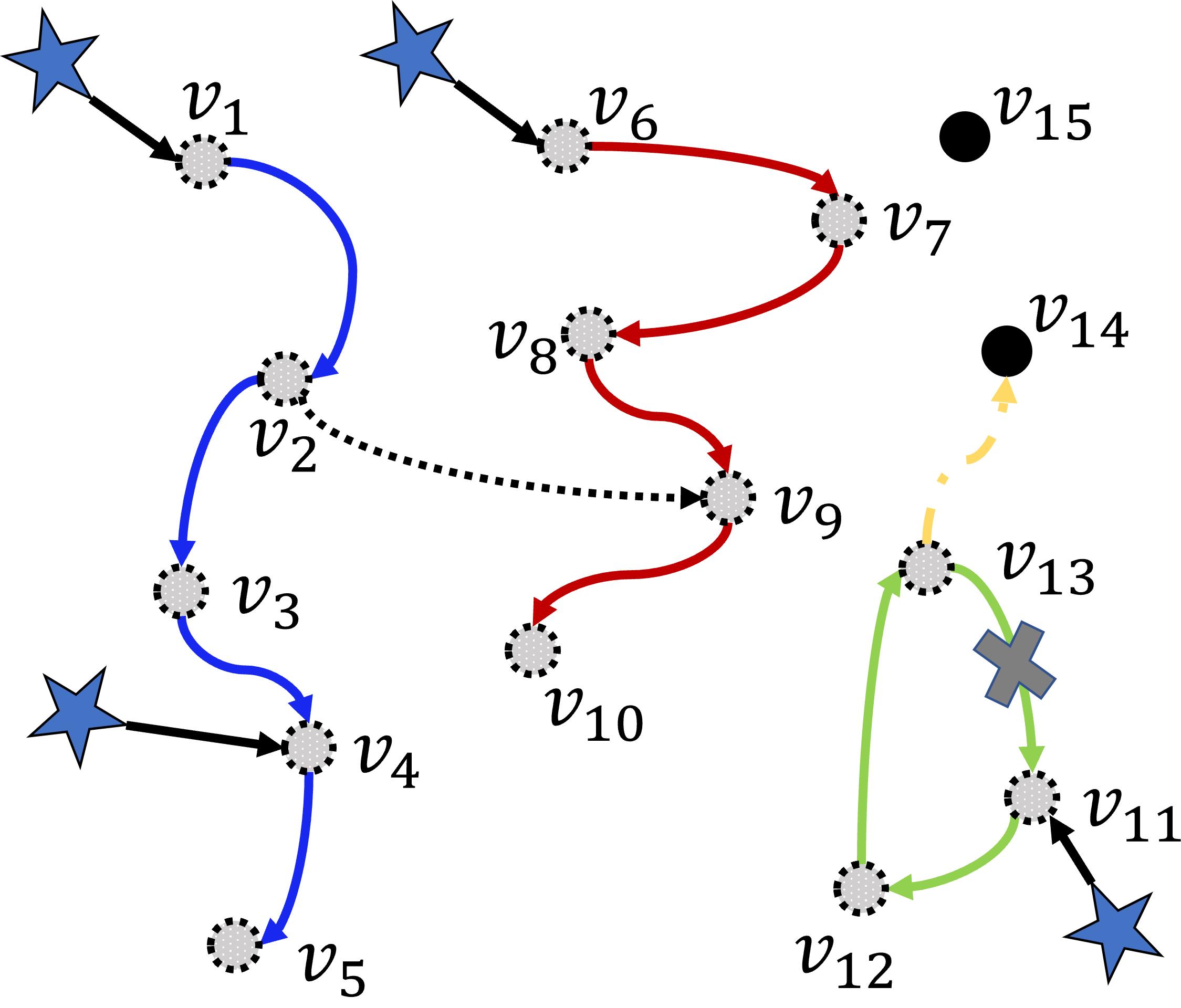}
		\caption{\label{Fig.sub.edcp5}}
	\end{subfigure}
 \hspace{2cm}
	\begin{subfigure}[t]{0.3\linewidth}
		\includegraphics[width=1\linewidth]{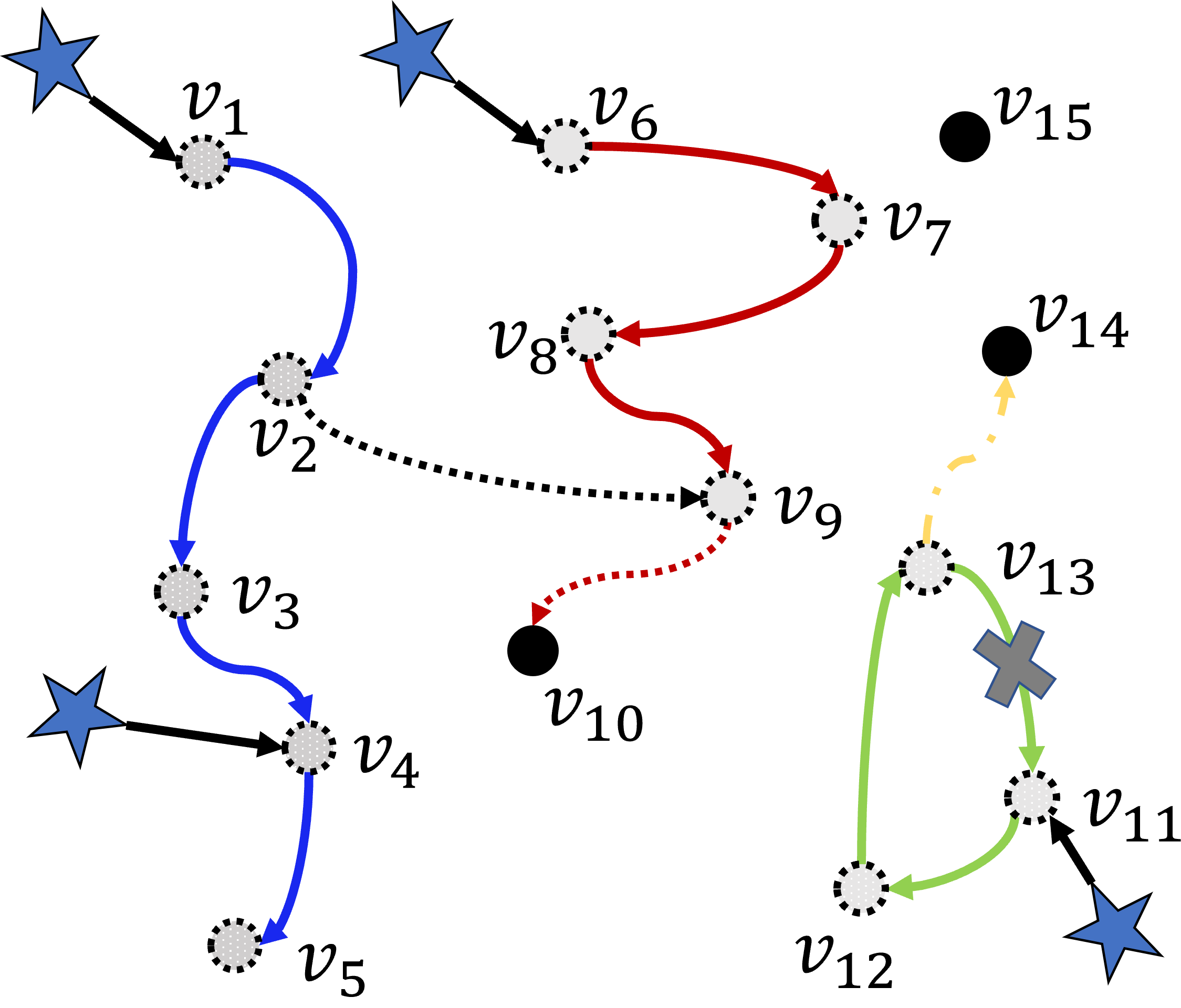}
		\caption{\label{Fig.sub.edcp6}}
	\end{subfigure}
	
	\caption[An example of the EDCP algorithm]
	{An example of the EDCP algorithm with $M = 4$ and $T = 12$. (a) The network has 15 nodes and 12 edges. (b) The EDCP algorithm starts with 4 paths and 1 cycle obtained by a certain controllability algorithm. 
		(c) Merge the cycle into paths. 
		(d) Assign the drive nodes in the paths until at least $T$ nodes are controllable. The input vertices represented as stars are connected to the first node of each sub-path. The nodes covered by sub-paths are selected to be controlled, and these nodes are highlighted in grey.
		(e) Limit the number of driver nodes to be $M$.
		(f) Decrease the number of nodes covered by sub-paths to $T$. }
	\label{Fig.edcp}
\end{figure}

It is easy to observe that the number of controllers may exceed $M$ by using the above method. We now discuss how to limit the number of driver nodes to $M$. For each path $\mathcal{p}_k \in \mathcal{P}$, the number of controlled nodes on it can be denoted as $|\mathcal{p}_k \setminus \mathcal{l}_k|$ and the number of driver nodes along with it is $d_k$. We could easily calculate the increased associate control cost $E' = \mathscr{E}(|\mathcal{p}_k \setminus \mathcal{l}_k|, d_k-1) - \mathscr{E}(|\mathcal{p}_k \setminus \mathcal{l}_k|, d_k)$ if one driver node on this path is to be released and the rest $d_k-1$ controllers are evenly distributed as much as possible to control the $|\mathcal{p}_k \setminus \mathcal{l}_k|$ nodes. We release the driver node from the path with the least increase in control cost, denote this path as $\mathcal{p}_m$, and re-allocate the $d_m-1$ driver nodes along this path. By repeating this procedure, the number of driver nodes is reduced to $M$.

Finally, we handle the case where the number of controllable nodes exceeds $T$. We could repetitively release one node from the longest sub-path until the number of nodes covered by the sub-paths exactly equals $T$. 

A summary of the algorithm is presented in Algorithm \ref{alg:EDCP Algorithm}, where lines 1 - 5 are for initialization, line 6 is for evenly dividing $T$ into $M$ parts, lines 7 - 18 are for assigning driver nodes to control at least $T$ nodes, lines 19 - 24 are for limiting the number of driver nodes to $M$ and lines 25 - 27 are for decreasing the number of nodes covered by sub-paths to $T$.

A simple example is presented in Fig. \ref{Fig.edcp} to illustrate the implementation of the EDCP algorithm. As shown in Fig. \ref{Fig.edcp}(a), the network has 15 nodes and 12 edges. The number of driver nodes and the number of nodes that need to be controlled are set to $M = 4$ and $T = 12$, respectively. After the MCFP algorithm, we obtain four paths and one cycle as shown in Fig. \ref{Fig.edcp}(b). Path ($v_1, v_2, v_3, v_4, v_5$) with the length 5 is highlighted in blue. Path ($v_6, v_7, v_8, v_9, v_{10}$) also with the length 5 is in red. Path $v_{14}$ and path $v_{15}$ have a length of 1. And cycle ($v_{11}, v_{12}, v_{13}, v_{11}$) in green is of length 3. The dashed edge $(v_2, v_9)$ does not belong to any path or cycle obtained by the MCFP algorithm.

% \section*{}

Now we use the EDCP algorithm to locate the driver nodes. As shown in Fig. \ref{Fig.edcp}(c), we first break the green cycle by removing the edge ($v_{13}, v_{11}$) and connect the cycle to the end node of the shortest path. A yellow dot-dash edge ($v_{14}, v_{11}$) is added. $T = 12$ is divided into $M = 4$ integers, yielding $Q = \{3,3,3,3\}$.

We then assign the driver nodes to the paths until at least $T$ nodes are controllable. External control inputs are shown as stars. As shown in Fig. \ref{Fig.edcp}(c), since the blue path has the longest undivided part with a length of 5. We set its first node ($v_1$) as a driver node and the first 3 nodes ($v_1$, $v_2$, and $v_3$) as controllable nodes, which are indicated by the color grey. Then the red path has the longest undivided part with a length of 5. We set its first node ($v_6$) as a driver node and the first 3 nodes ($v_6$, $v_7$, and $v_8$) as controllable nodes. The next driver node should be the first node of the green path ($v_{11}$) and the first 3 nodes ($v_{11}$, $v_{12}$, and $v_{13}$) in the green path become controllable. Now the blue path has the longest undivided part, which however has a length shorter than 3. We set its first node ($v_4$) as a driver node, and all nodes in such an undivided part become controllable. Finally, it would be the red path, where node $v_9$ is selected as a driver node and $v_9$ and $v_{10}$ are selected to be controlled.

Next, we reduce the number of driver nodes to $M = 4$. As shown in Fig. \ref{Fig.edcp}(e), when one driver node is removed, the blue and red paths have the same increased control cost. In this example, we randomly choose the red path and release one driver node from it. Thus the only driver node left in the red path is $v_6$.

Finally, we decrease the number of nodes covered by all sub-paths to $T = 12$. As shown in Fig. \ref{Fig.edcp}(f), the longest path is the red path with a length of 5. Remove node $v_{10}$ from the controlled node set. At last, we obtain 4 sub-paths namely ($v_1, v_2, v_3$), ($v_4, v_5$), ($v_6, v_7, v_8, v_9$) and ($v_{11}, v_{12}. v_{13}$). The first node of each sub-path is the driver node and the nodes covered by the sub-paths are selected as controlled nodes.

\subsection{Experimental Results}
We firstly apply both the eLPGM algorithm and the EDCP algorithm on ER \cite{Erdoes1960}, BA \cite{BARABASI2003} and some real-life networks with small or moderate sizes (CrystalID, Florida \cite{ulanowicz2012growth}, StMarks \cite{Baird1998}, Maspalomas \cite{Almunia1999}, C. elegans \cite{Chen2006}, Cons-Freq-Rev \cite{doi:10.1080/15427951.2009.10129177}, yeast \cite{Milo2002}, Cocomac \cite{Bakker2012}).

The comparisons between EDCP and various projected gradient descend methods in 100-node directed ER and BA networks, 
where the ER network has 300 edges and the BA network has 344 edges, are presented in Table \ref{Table.table_edvp_vs_pgm}. We set $M = 32$ and $t_f = 2$. 
Since there is no existing algorithm for comparisons under general cases of sufficient control, we compare $\text{eLPGM}$ versus LPGM under the special case where $C$ is given and only $B$ is optimized, and versus TPGME where $B$ is given and only $C$ is optimized. It can be seen that eLPGM steadily outperforms LPGM and TPGME, as eLPGM can optimize two matrices $B$ and $C$ simultaneously, whereas LPGM or TPGME is only capable of optimizing one of the matrices $B$ or $C$. For the same reason, although EDCP has much lower complexity, it sometimes slightly outperforms LPGM and TPGM. Another interesting observation is that EDCP may even perform better than eLPGM in rare cases. This may be partially due to the fact that the projected gradient method converges to a local minimum and requests a relatively large number of repetitions with different starting points. Current results were obtained by twenty-time repetitions. Simulations with more extensive restarting operations shall be carried out in our future studies. 
The observations we stated above remain valid in all real-life networks we have tested, as presented in Table \ref{table_edvp_vs_pgm_real}.

\begin{table}[H]
	\centering
 
	\caption{Comparison of minimum-cost sufficient control costs for EDCP and projected gradient descent methods on ER and BA networks $(M=32)$}
%   \resizebox{0.56\columnwidth}{!}{
  \setlength{\tabcolsep}{8pt} % Default value: 6pt
  \renewcommand{\arraystretch}{1.1} % Default value: 1
	\begin{tabular}{l|c|c|c|c|c}
 \toprule[1.5pt]
 &$\frac{T}{N}$ & EDCP & $\text{LPGM}$ \cite{Gao2018}& $\text{TPGME}$ \cite{Chen2020}& $\text{eLPGM}$ \\
		 \midrule[1pt]	ER & 0.4 & $4.76\mathrm{E} 02$& $4.00\mathrm{E} 02$ & $5.26\mathrm{E} 02$ & $3.72\mathrm{E} 02$ \\
		$N: 100$ & 0.5 & $7.04\mathrm{E} 02$& $7.30\mathrm{E} 02$ & $6.39\mathrm{E} 02$ & $5.88\mathrm{E} 02$ \\
		edges: $300$ & 0.6 & $7.90\mathrm{E} 02$ & $9.05\mathrm{E} 02$ & $8.86\mathrm{E} 02$ & $7.75\mathrm{E} 02$ \\
		& 0.7 & $1.11\mathrm{E} 03$& $1.20\mathrm{E} 03$ & $1.15\mathrm{E} 03$ & $9.91\mathrm{E} 02$ \\
		& 0.8 & $1.76\mathrm{E} 03$ & $1.79\mathrm{E} 03$ & $2.81\mathrm{E} 03$ & $1.56\mathrm{E} 03$ \\
		& 0.9 & $5.37\mathrm{E} 03$& $6.79\mathrm{E} 03$ & $6.09\mathrm{E} 03$ & $3.28\mathrm{E} 03$ \\	
		 \midrule[1pt]	BA & 0.4 & $2.26\mathrm{E} 02$ & $4.14\mathrm{E} 02$ & $4.42\mathrm{E} 02$ & $2.88\mathrm{E} 02$ \\
		$N:100$	& 0.5 & $4.12\mathrm{E} 02$& $5.96\mathrm{E} 02$ & $9.05\mathrm{E} 02$ & $5.20\mathrm{E} 02$ \\
		edges: $344$	& 0.6& $5.16\mathrm{E} 02$ & $1.12\mathrm{E} 03$ & $9.63\mathrm{E}02$ & $8.85\mathrm{E} 02$ \\
		& 0.7 & $8.44\mathrm{E} 02$ & $2.26\mathrm{E} 03$ & $1.65\mathrm{E} 03$ & $1.05\mathrm{E} 03$\\
		& 0.8 & $1.82\mathrm{E} 03$& $3.52\mathrm{E} 03$ & $4.01\mathrm{E} 03$ & $1.54\mathrm{E} 03$ \\
		& 0.9 & $7.40\mathrm{E} 03$ & $5.03\mathrm{E} 03$ & $5.69\mathrm{E} 03$ & $3.14\mathrm{E} 03$ \\
		\bottomrule[1.5pt]
	\end{tabular}
	\label{Table.table_edvp_vs_pgm}
\end{table}

\begin{table}[H]
	\centering
	\caption{Comparison of minimum-cost sufficient control costs for EDCP and projected gradient descent methods on real-life networks}
%	\resizebox{0.8\textwidth}{!}{
   \setlength{\tabcolsep}{8pt} % Default value: 6pt
  \renewcommand{\arraystretch}{1.1} % Default value: 1
 \begin{tabular}{l|c|c|c|c|c|c|c|c|c}
		\toprule[1.5pt] Networks &N &edges &$\frac{T}{N}$ &$T$ &$M$ &EDCP & $\text{LPGM}$ \cite{Gao2018} & $\text{TPGME}$ \cite{Chen2020} & $\text{eLPGM}$ \\
		\midrule[1pt] CrystalID \cite{ulanowicz2012growth} &24 &93 &0.6 &15 &10 &$0.48\mathrm{E} 02$&$0.44\mathrm{E} 02$ &$0.33\mathrm{E} 02$ &$0.28\mathrm{E} 02$\\
		 Florida \cite{ulanowicz2012growth} &128 &2106 &0.6 &77 &25 &$1.16\mathrm{E} 02$ &$0.97 \mathrm{E} 02$ &$1.16 \mathrm{E} 02$ &$0.45 \mathrm{E} 02$ \\
		 StMarks \cite{Baird1998} &54 &356 &0.7 &38 &25 &$2.74 \mathrm{E} 02$&$2.35\mathrm{E} 02$ &$3.13 \mathrm{E} 02$ &$2.14 \mathrm{E} 02$ \\
		 Maspalomas\cite{Almunia1999} &30 &87 &0.5 &15 &9 &$0.79 \mathrm{E} 02$&$0.98\mathrm{E} 02$ &$1.34 \mathrm{E} 02$ &$0.73 \mathrm{E} 02$ \\
		 C. elegans \cite{Chen2006}&297 &2345 &0.6 &179 &65 &$9.06 \mathrm{E} 02$&$8.67\mathrm{E} 02$ &$8.49 \mathrm{E} 02$ &$6.08 \mathrm{E} 02$ \\
	 Cons-Freq-Rev \cite{doi:10.1080/15427951.2009.10129177}&46 &879 &0.7 &33 &8 &$2.12 \mathrm{E} 02$&$1.34\text{E} 02$ &$2.12 \mathrm{E} 02$ &$1.14 \mathrm{E} 02$ \\
		 \bottomrule[1.5pt]
  	\end{tabular}
	\label{table_edvp_vs_pgm_real}
\end{table}

\begin{table}[H]
	\center
	\caption{Comparison of control costs on large-scale networks}
%	\resizebox{0.61\textwidth}{!}{
   \setlength{\tabcolsep}{8pt} % Default value: 6pt
  \renewcommand{\arraystretch}{1.1} % Default value: 1
		\begin{tabular}{l|c|c|c|c|c|c}
				
			\toprule[1.5pt] Networks & N & edges & $M$ & LPGM \cite{Gao2018}& MLCP \cite{Li2018}& EDCP\\
			\midrule[1pt] ER & 100 & 300 & 50 & $4.48\mathrm{E} 02$ & $8.75\mathrm{E} 02$ & $5.38\mathrm{E} 02$\\
			& 500 & 1493 & 300 & N.A. & $2.50\mathrm{E} 03$ & $1.45\mathrm{E} 03$\\
			& 1000 & 3004 & 550 & N.A. & $9.30\mathrm{E} 03$ & $3.69\mathrm{E} 03$\\
		\midrule[1pt] BA & 100 & 344 & 50 & $5.56\mathrm{E} 02$ & $6.49\mathrm{E} 02$ & $5.94\mathrm{E} 02$\\
			& 500 & 1549 & 300 & N.A. & $2.28\mathrm{E} 03$ & $1.42\mathrm{E} 03$\\
			& 1000 & 3021 & 550 & N.A. & $4.68\mathrm{E} 03$ & $3.75\mathrm{E} 03$\\
			 \midrule[1pt] yeast \cite{Milo2002} & 688 & 1079 & 570 & N.A. & $1.49\mathrm{E} 03$ & $1.50\mathrm{E} 03$\\
		 Cocomac \cite{Bakker2012}& 5336 & 36758 & 3900 & N.A. & $0.94\mathrm{E} 02$ & $0.94\mathrm{E} 02$\\
		\bottomrule[1.5pt]
		\end{tabular}
		\label{Table.table_edvp_vs_pgm_real_large}
\end{table}

Next, we test the performance of the EDCP algorithm on some large networks. Since there are no existing algorithms for large-scale minimum-cost sufficient control problems, we compare EDCP with MLCP \cite{Li2018} for the special case where the whole network is to be controlled. As shown in Table \ref{Table.table_edvp_vs_pgm_real_large}, EDCP's performance is typically better than MLCP. For those cases where networks are not too large and hence LPGM, a projected gradient descent method designed for achieving minimum-cost control of the whole network \cite{Li2018}, is applicable, EDCP performs comparably versus LPGM. As to why EDCP can usually outperform MLCP, it may be due to the fact that EDCP tries to make every control path be of nearly the same length while MLCP mainly focuses on minimizing the longest control path.

\section{Conclusion}
\label{sec:Conclusion}
In this paper, we investigated on the sufficient control problem in directed networks, including the sufficient controllability problem and the minimum-cost sufficient control problem. It was proved that the sufficient controllability problem is essentially a path cover problem. We rigorously proved that the path cover problem can be transformed into a minimum-cost flow problem. The MCFP algorithm was proposed to find the solution for the problem at polynomial complexity. It was observed that, in dense networks, having a few control sources could ensure the controllability of a sufficiently large portion of the network. For the minimum-cost sufficient control problem, which is an NP-hard problem, the eLPGM algorithm was proposed to solve the problem, obtaining suboptimal solutions on small- or medium-sized networks. For large-scale networks, the EDCP algorithm was proposed to tackle the problem at low complexity. Simulation results on synthetic and real-life networks evidently demonstrated the effectiveness of both the eLPGM and EDCP algorithms. Studies in this report have been based on LTI systems. Future work will be carried out to extend the results to complex networks with nonlinear dynamics.

\section*{Acknowledgment}
This work was partially supported by the Future Resilient Systems-Stage II (FRS-II) Programme at the Singapore-ETH Centre (SEC), funded by the National Research Foundation of Singapore (NRF), and by Ministry of Education of Singapore (Grant No. RG19/20). It was also partially supported by National Science Foundation of China (Grants No. 61876215 and 62002248), Beijing Academy of Artificial Intelligence (BAAI), and the China Postdoctoral Science Foundation under Grants No. 2019TQ0217 and 2020M673277.

% \section{Proof of Theorem \ref{theorem:gradient}}
% \label{sec:appendix}

 \bibliographystyle{elsarticle-num} 
 % \bibliography{cas-refs}
 \bibliography{IEEEabrv.bib,reference_sufficient_c.bib}


\appendix

\section{Proof of Theorem \ref{theorem_sufficient_controllability}}
\label{proof_sufficient_controllability}
% \begin{proof}
Briefly, the proof has two steps. By denoting the minimum-cost flow in the transformed network $D(V', E', b, c, s)$ as $f^*$ and the minimum-cost as $\sum_{e \in E'}f^*(e)c(e)$, 
the first step is to show that $M$ external controllers can achieve the structural controllability of $|\sum_{e \in E'}f^*(e)c(e)|$ nodes 
in network $D(V_{\text{A}}, E_{\text{A}})$. The second step is to demonstrate that the number of structural controllable nodes is no larger than $|\sum_{e \in E'}f^*(e)c(e)|$ with $M$ controllers.

We first prove that $M$ controllers can achieve the structural controllability of $|\sum_{e \in E'}f^*(e)c(e)|$ nodes in network $D(V_{\text{A}}, E_{\text{A}})$.
According to Lemmas \ref{lemma_structural_controllability} - \ref{lemma_target_controllability}, we need to show that there exist $M$ vertex-disjoint cacti covering $|\sum_{e \in E'} f^*(e)c(e)|$ nodes in $D(V_{\text{A}}, E_{\text{A}})$. Recall the network transformation discussed in Section \ref{subsec_Path_Cover_to_Minimum-cost_Flow}. We define a function $E^f = \Theta(f, E')$ where $f$ is a feasible flow in $D(V', E', b, c, s)$ and $E^f$ is the edge set consisting of those edges in $E'$ with positive flows, i.e., $E^f = \{e: f(e) = 1, e \in E'\}$. Since $s(v_s) = - s(v_t) = M$ and $s(v) = 0$, $\forall v \in V' \setminus V_\text{s, t}$, there are $M$ paths from $v_s$ to $v_t$ and an unspecified number of cycles, each of which is a sequence of nodes in $V^f$ connected by edges in $E^f$. These paths and cycles share no common nodes in $V' \setminus V_\text{s, t}$ as $b(e) = 1$, $\forall e \in E'$. Then, we denote the vertex set consisting of all nodes that are incident to at least 
one edge in $E^f$ by $V^f = \{v_i: (v_i, v_j) \in E^f \operatorname{or} \ \ (v_k, v_i) \in E^f \}$. 
If a node $v_i^{\text{in}} \in V_\text{A}^\text{in}$ satisfies that $v_i^{\text{in}} \in V^f$, then its corresponding vertex $v_i^{\text{out}} \in V_\text{A}^\text{out}$ also satisfies that $v_i^{\text{out}} \in V^f$ and vice versa. This is because 
\begin{equation}
\label{eq_supply_in}
  s(v_i^{\text{in}}) = 0 = \sum_{v_j \in V'}f(v_i^{\text{in}}, v_j) - \sum_{v_k \in V'}f(v_k, v_i^{\text{in}}),
\end{equation}
and 
\begin{equation}
\label{eq_supply_out}
  s(v_i^{\text{out}}) = 0 = \sum_{v_j \in V'}f(v_i^{\text{out}}, v_j) - \sum_{v_k \in V'}f(v_k, v_i^{\text{out}}).
\end{equation}
As $(v_i^{\text{in}}, v_i^{\text{out}})$ is the only incoming (outgoing) edge to $v_i^{in}$ (from $v_i^{out}$), if $v_i^{\text{in}} \in V^f$ or $v_i^{\text{out}} \in V^f$, then we must have $f(v_i^{in}, v_i^{out}) > 0$
according to (\ref{eq_supply_in}) and (\ref{eq_supply_out}). The edge set $E^{f^*}$ for the minimum-cost flow $f^*$ can be defined as $E^{f^*} = \Theta(f^*, E') = \{e: f^*(e) = 1, e \in E'\}$. And vertex set $V^{f^*}$ is denoted by $V^{f^*} = \{v_i: (v_i, v_j) \in E^{f^*} \operatorname{or} \ \ (v_k, v_i) \in E^{f^*} \}$. The $M$ paths and an unspecified number of cycles consisting of nodes in $V^{f^*}$ connected by edges in $E^{f^*}$ can be mapped into corresponding $M$ vertex-disjoint paths and an unspecified number of cycles in the original network $D(V_{\text{A}}, E_{\text{A}})$. Such paths and cycles form the vertex-disjoint path set $\mathcal{P}$ and cycle set $\mathcal{C}$, respectively, where $|\mathcal{P}| = M$. 
Since $c(e) = -1$, $\forall e \in E'_{\text{inner}}$ where $E'_{\text{inner}} = \{(v_i^{\text{in}}, v_i^{\text{out}}): v_i \in V_{\text{A}}\}$ and $c(e) = 0$, $\forall e \in E' \setminus E'_{\text{inner}}$, we have $1/2 |V^{f^*} \setminus V_{\text{s,t}}| = |\operatorname{cover}(\mathcal{P} \cup \mathcal{C})| = |\sum_{e \in E'}f^*(e)c(e)|$.

Next, we shall show that no more than $|\sum_{e \in E'}f^*(e)c(e)|$ nodes in network $D(V_{\text{A}}, E_{\text{A}})$ can be structurally controllable with $M$ controllers. This can be proved by contradiction. We assume that $M$ controllers can satisfy $T'$ nodes' structural controllability in $D(V_{\text{A}}, E_{\text{A}})$ where $T' > |\sum_{e \in E'}f^*(e)c(e)|$. This implies $-T' < \sum_{e \in E'}f^*(e)c(e)$. Based on the network transformation illustrated in Section \ref{subsec_Path_Cover_to_Minimum-cost_Flow} and Lemma \ref{lemma_target_controllability}, having $M$ controllers satisfying $T'$ nodes' structural controllability in $D(V_{\text{A}}, E_{\text{A}})$ implies that there are $M$ vertex-disjoint paths and an unspecified number of cycles covering $2T'$ nodes belonging to $V' \setminus V_\text{s,t}$ in $D(V', E', b, c, s)$. 
For every node $v_i$ covered by these paths or cycles, we have $f(v_i^{\text{in}}, v_i^{\text{out}}) = 1$ and $(v_i^{\text{in}}, v_i^{\text{out}}) \in E'_\text{I}$. Since $c(e) = -1$, $\forall e \in E'_\text{I}$ and $c(e) = 0$, $\forall e \in E' \setminus E'_\text{I}$, the total cost of the flow $f$ over all edges in $E'$ equals $-T'$ which is smaller than the total cost of the minimum cost flow $\sum_{e \in E'}f^*(e)c(e)$. Therefore, we can conclude that the assumption that $-T' < \sum_{e \in E'}f^*(e)c(e)$ leads to a contradiction.
\qed
% \end{proof}

\section{Proof of Theorem \ref{theorem_matrix_bc_derivative}}
\label{proof_matrix_bc_derivative}
Before obtaining the gradient information of $B$ and $C$, we first review the following matrix derivative and chain rule. 

\begin{lemma}
\label{lemma_matrix_basics}
\cite{Petersen2012}
  For an independent matrix $X$ with $x_{ij}$ or $[X]_{ij}$ as its $(i,j)^{\text{th}}$ entry, denote that $\delta_{i j} = 1$ if and only if $i = j$; and $\delta_{i j} = 0$ otherwise. We have:\\
a) $\frac{\partial x_{k l}}{\partial x_{i j}}=\delta_{k i} \cdot \delta_{l j}$, $\frac{\partial x_{k l}^{\text{T}}}{\partial x_{i j}}=\delta_{k j} \cdot \delta_{l i}$.\\
b) $\frac{\partial x_{k l}^{-1}}{\partial x_{i j}}=-x_{k i}^{-1} \cdot x_{j l}^{-1}$.\\
c) $\delta_{k i} \cdot \delta_{i j}=\delta_{k j}$.\\
d) $x_{i j} \cdot \delta_{j a}=x_{i a}$, $\delta_{i a} \cdot x_{i j}=\left[x^{\text{T}}\right]_{j i} \cdot \delta_{i a}=\left[x^{\text{T}}\right]_{j a}=x_{a j}$.\\
e) For compatible matrices $X$, $Y$, $Z$ such that $X = YZ$, then $x_{i j}=\sum_{k} y_{i k} z_{k j}$. It can be written as $x_{i j}=y_{i k} \cdot z_{k j}$.
\end{lemma}
 
\begin{lemma}
    \label{lemma:chain_rule}
\cite{Petersen2012}
{\it chain rule}: Let $Z = p(X)$, the derivative of the function $q(Z)$ with respect to $X$ is 
\begin{equation}
    \frac{\partial q(Z)}{\partial X} = \frac{\partial q(Z)}{x_{ij}} = \sum_{k=1}^{M} \sum_{l=1}^{N} \frac{\partial q(Z)}{\partial z_{kl}}\frac{\partial z_{kl}}{\partial x_{ij}}.
\end{equation}
\end{lemma}

According to Lemmas \ref{lemma_matrix_basics} - \ref{lemma:chain_rule}, by varying control cost $J$ in (\ref{E.optimalsufficientcontrol}) with respect to $B$, we find that 

% \begin{equation}
	\begin{align*}
			\frac{\partial J(B, C)}{\partial b_{i j}} =&\frac{\partial \operatorname{tr}\left(C^{\text{T}}\left(C W C^{\text{T}}\right)^{-1} C e^{A t_{f}} e^{A^{\text{T}} t_{f}}\right)}{\partial b_{i j}} \\
			=& \frac{\partial \operatorname{tr}\left(C^{\text{T}}\left(C W C^{\text{T}}\right)^{-1} C e^{A t_{f}} e^{A^{\text{T}} t_{f}}\right)}{\partial\left[C W C^{\text{T}}\right]_{k l}} \cdot \frac{\partial\left[C W C^{\text{T}}\right]_{k l}}{\partial b_{i j}} \\
			=&-\left[\left(C W C^{\text{T}}\right)^{-\text{T}} C e^{A t_{f}} e^{A^{\text{T}} t_{f}} C^{\text{T}}\left(C W C^{\text{T}}\right)^{-\text{T}}\right]_{k l} 
			 \cdot \frac{\partial\left([C]_{k m} \int_{0}^{t_{f}}\left[e^{A t}\right]_{m n}[B]_{n q}\left[B^{\text{T}}\right]_{q w}\left[e^{A^{\text{T}} t}\right]_{w s} d t\left[C^{\text{T}}\right]_{s l}\right)}{\partial b_{i j}} \\
			=&-\left[\left(C W C^{\text{T}}\right)^{-\text{T}} C e^{A t_{f}} e^{A^{\text{T}} t_{f}} C^{\text{T}}\left(C W C^{\text{T}}\right)^{-\text{T}}\right]_{k l}\\
   &\cdot\left([C]_{k m} \cdot \int_{0}^{t_{f}}\left[e^{A t}\right]_{m n} \cdot\left[\delta_{n i} \cdot \delta_{q j} \cdot\left[B^{\text{T}}\right]_{q w} +[B]_{n q} \cdot \delta_{q j} \cdot \delta_{w i}\right] \cdot\left[e^{A^{\text{T}} t}\right]_{w s} d t \cdot\left[C^{\text{T}}\right]_{s l}\right) \\
			=&-\left[\left(C W C^{\text{T}}\right)^{-\text{T}} C e^{A t_{f}} e^{A^{\text{T}} t_{f}} C^{\text{T}}\left(C W C^{\text{T}}\right)^{-\text{T}}\right]_{k l} \cdot[C]_{k m} \cdot \int_{0}^{t_{f}}\left[e^{A t}\right]_{m i} \cdot\left[B^{\text{T}}\right]_{j w} \cdot\left[e^{A^{\text{T}} t}\right]_{w s} d t \cdot\left[C^{\text{T}}\right]_{s l} \\
			=&-\left[\left(C W C^{\text{T}}\right)^{-\text{T}} C e^{A t_{f}} e^{A^{\text{T}} t_{f}} C^{\text{T}}\left(C W C^{\text{T}}\right)^{-\text{T}}\right]_{k l} \cdot[C]_{k m} \cdot \int_{0}^{t_{f}}\left[e^{A t}\right]_{m n} \cdot[B]_{n j} \cdot\left[e^{A^{\text{T}} t}\right]_{i s} d t \cdot\left[C^{\text{T}}\right]_{s l} \\
			=&-2 \cdot \int_{0}^{t_{f}}\left[e^{A^{\text{T}} t}\right]_{i m} \cdot\left[C^{\text{T}}\right]_{m k} \cdot\left[\left(C W C^{\text{T}}\right)^{-\text{T}} C e^{A t_{f}} e^{A^{\text{T}} t_{f}} C^{\text{T}}\left(C W C^{\text{T}}\right)^{-\text{T}}\right]_{k l} 
			\cdot[C]_{l s} \cdot\left[e^{A t}\right]_{s w} \cdot[B]_{w j} d t \\
			=&-2\left[\int_{0}^{t_{f}} e^{A^{\text{T}} t} C^{\text{T}}\left(C W C^{\text{T}}\right)^{-\text{T}} C e^{A t_{f}} e^{A^{\text{T}} t_{f}} C^{\text{T}}\left(C W C^{\text{T}}\right)^{-\text{T}} C e^{A t} B d t\right]_{i j}
	\end{align*}	
% \end{equation}		
where $k, l, m, n, i, q, j, w, s, l$ are the indexes of elements in the matrix. It can be derived that
\begin{equation}
		 \frac{\partial J(B, C)}{\partial B} = -2\left[\int_{0}^{t_{f}} e^{A^{\text{T}} t} C^{\text{T}}\left(C W C^{\text{T}}\right)^{-\text{T}} C e^{A t_{f}} e^{A^{\text{T}} t_{f}} C^{\text{T}}\left(C W C^{\text{T}}\right)^{-\text{T}} C e^{A t} B d t\right].
\end{equation}

And by varying control cost $J$ in (\ref{E.optimalsufficientcontrol}) with respect to $C$, we have

% \begin{equation}
	\begin{align*}
	\frac{\partial J(B, C)}{\partial c_{j i}} =&\frac{\partial \operatorname{tr}\left(C^{\text{T}}\left(C W C^{\text{T}}\right)^{-1} C e^{A t_{f}} e^{A^{\text{T}} t_{f}}\right)}{\partial c_{j i}}\\
	= & \frac{\partial C^{\text{T}}_{ab}}{\partial c_{ji}} \cdot \left[ CWC^{\text{T}} \right]_{bc}^{-1} \cdot c_{cd} \cdot [e^{A t_{f}} e^{A^{\text{T}} t_{f}}]_{da} + c_{ba} \cdot \frac{\partial \left[ CWC^{\text{T}} \right]_{bc}^{-1}}{\partial \left[ CWC^{\text{T}} \right]_{ef}} \cdot {\frac{\partial  \left[ CWC^{\text{T}} \right]_{ef}   }{\partial c_{ji}}} \cdot c_{cd}\cdot [e^{A t_{f}} e^{A^{\text{T}} t_{f}}]_{da} \\
&+ c_{ba} \cdot \left[ CWC^{\text{T}} \right]_{bc}^{-1} \cdot {\frac{\partial c_{cd} }{\partial c_{ji}}} \cdot[e^{A t_{f}} e^{A^{\text{T}} t_{f}}]_{da}\\
	= & \delta_{ai} \cdot \delta_{bj} \cdot [CWC^{\text{T}}]^{-1}_{bc} \cdot c_{cd} \cdot [e^{A t_{f}} e^{A^{\text{T}} t_{f}}]_{da}-
 c_{ba} \cdot   [CWC^{\text{T}}]^{-1}_{be} \cdot [CWC^{\text{T}}]^{-1}_{fc} \\
 &\cdot \left\{\delta_{ej} \cdot \delta_{gi} \cdot w_{gh} \cdot c_{fh} +c_{eg} \cdot w_{gh} \cdot \delta_{hi} \cdot \delta_{fj}\right\} \cdot c_{cd} \cdot [e^{A t_{f}} e^{A^{\text{T}} t_{f}}]_{da}
 +c_{ba} \cdot [CWC^{\text{T}}]^{-1}_{bc} \cdot \delta_{cj} \cdot \delta_{di}\left[e^{A t_{f}} e^{A^{\text{T}} t_{f}}\right]_{da}\\
	= &\delta_{ai} \cdot \delta_{bj} \cdot [CWC^{\text{T}}]^{-1}_{bc} \cdot c_{cd} \cdot [e^{A t_{f}} e^{A^{\text{T}} t_{f}}]_{da}-c_{ba} \cdot   [CWC^{\text{T}}]^{-1}_{be} \cdot [CWC^{\text{T}}]^{-1}_{fc} \cdot \delta_{ej} \cdot \delta_{gi} \cdot w_{gh} \cdot c_{fh} \cdot c_{cd}\cdot [e^{A t_{f}} e^{A^{\text{T}} t_{f}}]_{da} \\
	&-c_{ba} \cdot ( [CWC^{\text{T}}]^{-1}_{be} \cdot [CWC^{\text{T}}]^{-1}_{fc}) \cdot c_{eg} \cdot w_{gh} \cdot \delta_{hi}  \cdot \delta_{fj} \cdot c_{cd}\cdot [e^{A t_{f}} e^{A^{\text{T}} t_{f}}]_{da} +c_{ba} \cdot [CWC^{\text{T}}]^{-1}_{bc} \cdot  \delta_{cj} \cdot \delta_{di} \cdot \left[e^{A t_{f}} e^{A^{\text{T}} t_{f}}\right]_{da}\\
	= & \delta_{ai} \cdot [e^{A t_{f}} e^{A^{\text{T}} t_{f}}]_{da} \cdot c_{dc} \cdot [CWC^{\text{T}}]^{-1}_{cb} \cdot \delta_{bj}- w_{ih} \cdot  c_{ba} \cdot [CWC^{\text{T}}]^{-1}_{fc} \cdot c_{cd} \cdot [e^{A t_{f}} e^{A^{\text{T}} t_{f}}]_{da} \cdot c_{ba}  [CWC^{\text{T}}]^{-1}_{be}  \cdot \delta_{ej} \\
	&-w_{ig} \cdot  c_{eg}  [CWC^{\text{T}}]^{\text{-T}}_{eb} \cdot c_{ba} \cdot [e^{A t_{f}} e^{A^{\text{T}} t_{f}}]^{\text{T}}_{ad} \cdot c_{cd} \cdot [CWC^{\text{T}}]^{\text{-T}}_{cf}  \cdot \delta_{fj} +\delta_{di} \cdot  [e^{A t_{f}} e^{A^{\text{T}} t_{f}}]_{da} \cdot c_{ba} \cdot [CWC^{\text{T}}]^{-1}_{bc} \cdot \delta_{cj} \\
		= &\left[e^{A t_{f}} e^{A^{\text{T}} t_{f}} C^{\text{T}} [CWC^{\text{T}}]^{-1}\right]_{ij} -\left[W C^{\text{T}} [CWC^{\text{T}}]^{-1} C e^{A t_{f}} e^{A^{\text{T}} t_{f}} C^{\text{T}} [CWC^{\text{T}}]^{-1}\right]_{ij} \\
  &-\left[W C^{\text{T}} [CWC^{\text{T}}]^{-1} C e^{A t_{f}} e^{A^{\text{T}} t_{f}} C^{\text{T}} [CWC^{\text{T}}]^{-1}\right]_{ij} +\left[e^{A t_{f}} e^{A^{\text{T}} t_{f}} C^{\text{T}} [CWC^{\text{T}}]^{-1}\right]_{ij}\\
	 =&-2 [W C^{\text{T}}\left(C W C^{\text{T}}\right)^{-1} C e^{A t_{f}} e^{A^{\text{T}} t_{f}} C^{\text{T}}\left(C W C^{\text{T}}\right)^{-1}]_{ij}
	+2 [e^{A t_{f}} e^{A^{\text{T}} t_{f}} C^{\text{T}}\left(C W C^{\text{T}}\right)^{-1}]_{ij},
	\end{align*}
% \end{equation}
where $a, b, c, d, e, f, g, h, i, j, k$ are the indexes of elements in the matrix. Thus, we obtain 
\begin{equation}
		\begin{aligned}
			 \frac{\partial J(B, C)}{\partial C} = 
   -2\left(C W C^{\text{T}}\right)^{-1}C e^{A t_{f}} e^{A^{\text{T}} t_{f}}C^{\text{T}}\left(C W C^{\text{T}}\right)^{-1}CW^{\text{T}} + 2\left(C W C^{\text{T}}\right)^{-1}Ce^{A t_{f}} e^{A^{\text{T}} t_{f}}.
   % -2 W C^{\text{T}}\left(C W C^{\text{T}}\right)^{-1} C e^{A t_{f}} e^{A^{\text{T}} t_{f}} C^{\text{T}}\left(C W C^{\text{T}}\right)^{-1}
			% +2 e^{A t_{f}} e^{A^{\text{T}} t_{f}} C^{\text{T}}\left(C W C^{\text{T}}\right)^{-1}.
		\end{aligned}
\end{equation}

\qed

%% References


% \bibliographystyle{elsarticle-num}
% \bibliography{<your-bib-database>}






% \end{thebibliography}

\end{document}